\documentclass[reprint,floatfix,prb,twocolumn,showkeys,showpacs,superscriptaddress,notitlepage]{revtex4-1}
\usepackage{braket}
\usepackage{dsfont}
\usepackage{amsmath,amssymb,times,color,graphicx,multirow,amsfonts,color}
\usepackage[colorlinks=true,citecolor=blue,linkcolor=blue]{hyperref}
\usepackage{physics,bm,here}
\usepackage{silence}

\renewcommand{\ket}[1]{\ensuremath{\vert#1\rangle}}
\renewcommand{\bra}[1]{\ensuremath{\langle#1\vert}}
\renewcommand{\va}{{\vec a}}
\renewcommand{\vr}{{\vec r}}
\newcommand{\vR}{{\vec R}}
\newcommand{\vx}{{\vec x}}
\newcommand{\ve}{{\vec e}}

\newcommand{\vP}{\vec{\mathfrak P}}
\DeclareFontEncoding{LS1}{}{}
\DeclareFontSubstitution{LS1}{stix}{m}{n}

\begin{document}
\title{Bulk-and-edge to corner correspondence} 
\author{Luka Trifunovic} 
\email{luka.trifunovic@uzh.ch}
\affiliation{Department of Physics, University of Zurich, Winterthurerstrasse 190, 8057 Zurich, Switzerland}
\begin{abstract}
	We show that two-dimensional band insulators, with vanishing bulk
	polarization, obey bulk-and-edge to corner charge correspondence,
	stating that the knowledge of the bulk and the two corresponding ribbon
	band structures uniquely determines a fractional part of the corner
	charge irrespective of the corner termination. Moreover, physical
	observables related to macroscopic charge density of a
	terminated crystal can be obtained by representing the crystal as
	collection of polarized edge regions with polarizations $\vec
	P^\text{edge}_\alpha$, where the integer $\alpha$ enumerates the edges.
	We introduce a particular manner of cutting a crystal, dubbed ``Wannier
	cut'', which allows us to compute $\vec P^\text{edge}_\alpha$. We find
	that $\vec P^\text{edge}_\alpha$ consists of two pieces: the bulk piece
	expressed via quadrupole tensor of the bulk Wannier functions' charge
	density and the edge piece corresponding to the Wannier edge
	polarization---the polarization of the edge subsystem obtained by
	Wannier cut. For a crystal with $n$ edges, out of $2n$ independent
	components of $\vec P^\text{edge}_\alpha$, only $2n-1$ are independent
	of the choice of Wannier cut and correspond to physical observables:
	corner charges and edge dipoles.
\end{abstract}

\date{\today}
\maketitle
\section{Introduction}
While the bulk description of solid-state materials is generally available, the
description close to the material's boundaries (termination) is often not
accessible. For this reason, a particularly important role for material science
is played by \textit{bulk quantities}---they depend only on the
material's bulk although they predict a certain quantity that can be measured
once a boundary is introduced. In other words, the sole existence of the bulk
quantities requires some form of bulk-boundary correspondence. To name a few
examples, the bulk electrical polarization of an insulator predicts a fractional
part of the end
charge~\cite{resta2007,king-smith1993,vanderbilt1993,resta1994,resta1999}, the
bulk orbital magnetization~\cite{thonhauser2005,shi2007} predicts persistent
current circulating along the boundary, bulk geometric orbital
magnetization~\cite{trifunovic2019b} predicts a fractional part of the
time-averaged edge current circulating along the boundary of a periodically,
adiabatically driven insulator, and the bulk magnetoelectric polarizability of
a three-dimensional insulator predicts a fractional part of the surface
charge density resulting from the application of an external magnetic
field.~\cite{qi2008,essin2009}

In recent years,~\cite{kitaev2009,schnyder2009} the term bulk-boundary
correspondence is almost exclusively used in the context of topological
phenomena. In that more strict sense, the bulk-boundary correspondence assumes
that the bulk quantity is topological invariant; hence, the boundary quantity
is quantized. Notable examples include quantum (spin) Hall effect where Chern
number (Kane-Mele invariant~\cite{kane2005b}) predicts quantized (spin)
Hall conductance,~\cite{thouless1982,kane2005,bernevig2006}, and $\mathbb{Z}_2$
invariant predicting quantized zero-energy conductance of the Kitaev
chain.~\cite{kitaev2001,alicea2011,lutchyn2010,mourik2012} In this work, such
correspondence is referred to as topological bulk-boundary
correspondence.~\cite{schnyder2009,shiozaki2018,trifunovic2019,roberts2020,trifunovic2020}
In certain cases bulk-boundary correspondence can be enriched by the attribute
``topological'' in the presence of certain symmetries: The bulk polarization
and a fractional part of the end charge become quantized in the presence of
inversion symmetry, the bulk geometric orbital magnetization and a fractional part of
the time-averaged edge current are quantized in the presence of inversion or
fourfold rotation symmetry,~\cite{trifunovic2019b} and similarly, the
magnetoelectric polarizability and the associated boundary quantity are
quantized in the presence of time-reversal or inversion symmetry.~\cite{qi2008}
On the other hand, no symmetry quantizes the bulk orbital magnetization. It may
be of interest to ask a reverse question: In which cases
a topological bulk-boundary correspondence can be extended to its unquantized
version? This work deals with one example where such an extension is not
possible---the bulk quadrupole moment and the corner charge. Namely, in the
presence of fourfold rotation symmetry, the bulk quadrupole moment is
topological invariant and predicts the quantized corner
charge,~\cite{benalcazar2017} whereas in the absence of this symmetry
constraint it is not possible to predict the corner charge without specifying
the edge terminations.~\cite{ono2019} In this work, we show that instead of bulk
to boundary correspondence, a bulk-and-edge to corner correspondence can be
formulated.
\begin{figure}[t]
	\centering
	\includegraphics[width=0.6\columnwidth]{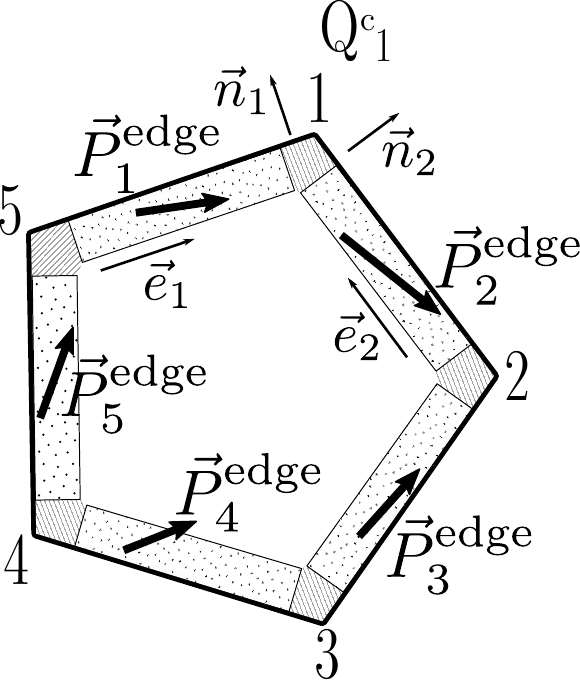}
	\caption{Illustration of a crystal with boundary. To obtain a
		fractional part of each of the five corners (hatched regions),
		the bulk and the five edge terminations (dotted regions) need to
		be specified, while the termination around the corners need not
		be specified. The edges run along the lattice vectors $\vec
		e_\alpha$, with the unit normal vectors $\vec n_\alpha$ pointing
		outward, $\alpha=1,\dots,5$. The corner charge, where the
		edges along $\vec e_\alpha$ and $\vec e_{\alpha+1}$ meet, is
		denoted by $Q_\alpha^\text{c}$. The macroscopic charge density
		of a crystal can be seen to be generated by a collection of
		polarized edge regions (dotted) with polarizations $\vec
		P^\text{edge}_\alpha$. The charge neutrality and vanishing
		bulk polarization are assumed; hence, $\sum_\alpha
		Q^\text{c}_\alpha=0$, whereas $\sum_\alpha\vec
		P^\text{edge}_\alpha$ need not vanish.}
	\label{fig:1}
\end{figure}

In 2015, Zhou, Rabe, and Vanderbilt~\cite{zhou2015} proposed that for band
insulators, a fractional part of the corner charge $Q^\text{c}$ can be
computed from the knowledge of the bulk and the two corresponding ribbon band
structures via the following relation:
\begin{align}
	Q^\text{c}&=P^\text{edge}_x+P^\text{edge}_y\mod e,
	\label{eq:Qcwrong}
\end{align}
where $P^\text{edge}_x$ ($P^\text{edge}_y$) is the $x$ component ($y$ component) of
the edge polarization for the edge along the $x$ ($y$) direction. These
authors defined the edge polarization in terms of so-called
maximally localized hybrid Wannier functions,~\cite{marzari2012} and verified
the relation~(\ref{eq:Qcwrong}) using two tight-binding models.~\cite{zhou2015}
One year later, in their pioneering work, Benalcazar, Bernevig, and Hughes
proposed the model~\cite{benalcazar2017} that in the presence of fourfold
rotation symmetry~\cite{benalcazar2019} exhibits quantized corner charge that
is given by topological invariant dubbed ``bulk quadrupole moment'' $q_{xy}$,
\begin{align}
	q_{xy}&=Q^\text{c}-P^\text{edge}_x-P^\text{edge}_y\mod e,
	\label{eq:2}
\end{align}
with $P^\text{edge}_{x,y}$ defined in the same manner as in
Eq.~(\ref{eq:Qcwrong}).  The nonvanishing value of
``$Q^\text{c}-P^\text{edge}_x-P^\text{edge}_y$'' proves that previously
proposed relation~(\ref{eq:Qcwrong}) cannot hold in general. Note that fourfold
rotation symmetry forces the relation $P^\text{edge}_x=-P^\text{edge}_y\mod e$ to
hold, and hence the $q_{xy}=Q^\text{c}\mod e$ which expresses topological bulk-boundary
correspondence.  Subsequent works~\cite{wheeler2019,kang2019} by two
independent groups proposed an expression that was meant to predict $q_{xy}$ in
the absence of the symmetry constraints, using the bulk Hamiltonian as its sole
input. These findings were supported by calculations on several tight-binding
models.~\cite{wheeler2019,kang2019} Shortly after, Ono, Watanabe, and the
present author provided counterexamples showing that the proposed expression
does not hold in general. In this work, we show that there exists no unique value
for $q_{xy}$, since the edge polarizations in Eq.~(\ref{eq:2}) are not uniquely
defined quantities.

This work considers two-dimensional band insulators with a well-defined corner
charge, which is the case when not only the bulk but also the boundary is
gapped and the edge charge density vanishes. Figure~\ref{fig:1} shows one
example of a terminated crystal, with the index $\alpha$ enumerating the
corners. Two edges, along lattice vectors $\vec e_\alpha$ and
$\vec e_{\alpha+1}$ with the corresponding unit normal vectors $\vec n_\alpha$
and $\vec n_{\alpha+1}$, meet at the corner with the index $\alpha$. The main
result of this work is finding that the physical observables related to
macroscopic~\cite{jackson1999} charge density $\rho^\text{macro}$ of such
terminated crystal---corner charges and edge dipoles---can be obtained by
representing the crystal as a collection of edge regions with polarizations
$\vec P^\text{edge}_\alpha$; see Fig.~\ref{fig:1}. We find that the edge
polarizations $\vec P^\text{edge}_\alpha$ consist of two pieces,
\begin{align}
	\vec P^\text{edge}_\alpha=& L_\alpha\hat q\cdot\vec n_{\alpha}/2+\vP^\text{edge}_\alpha,
	\label{eq:Pedge}
\end{align}
with $L_\alpha=\vert\vec e_\alpha\vert$ being the shortest repeated length
along the $\alpha$ direction. The quadrupole tensor density $\hat q$ (bulk
piece) is defined as the quadrupole tensor of the charge density of the
bulk Wannier functions divided by the area of the unit cell. The Wannier
edge polarization $\vP^{\text{edge}}_\alpha$ is obtained from the corresponding
ribbon band structure by performing a ``Wannier cut``; see Sec.~\ref{sec:Pedge}
for the precise definition. For a crystal in with $n$ edges, there are $n$ edge
polarizations with $2n$ independent components; out of those $2n-1$ are
independent of the choice of Wannier cut (see Sec.~\ref{ssec:discussion}).

The result~(\ref{eq:Pedge}) turns the bulk-boundary correspondence for
electrical polarization~\cite{king-smith1993} into bulk-and-edge to corner
correspondence
\begin{align}
	Q_\alpha^\text{c}=& \frac{L_{\alpha+1}}{A_\text{cell}}\vec P^\text{edge}_\alpha\cdot\vec n_{\alpha+1}+\frac{L_{\alpha}}{A_\text{cell}}\vec P^\text{edge}_{\alpha+1}\cdot\vec n_{\alpha}\mod e ,
	\label{eq:bcc}
\end{align}
where $A_\text{cell}=\vert\vec e_\alpha\times\vec e_{\alpha+1}\vert$ is the
area of the unit cell defined by the corresponding corner.

The remaining of the article is organized as follows. In Sec.~\ref{sec:pre}, we
review the modern theory of electrical polarizations and define corner charges and
edge dipoles. Section~\ref{sec:main} contains the main results of our work;
therein we formulate and prove bulk-and-edge to corner charge correspondence
and introduce the notion of Wannier cut and Wannier edge polarization. Three
simple tight-binding models that illustrate the procedure described in
Sec.~\ref{sec:main} can be found in Sec.~\ref{sec:examples}. We conclude in
Sec.~\ref{sec:conclusions}.

\section{Preliminaries}\label{sec:pre}
We start by reviewing the distinction between microscopic and macroscopic
charge density of a crystal. Section~\ref{sec:modth} reviews the modern theory
of electric polarization of band insulators and the corresponding bulk-boundary
correspondence.~\cite{resta2007,king-smith1993}

\subsection{Macroscopic charge density. Corner charge and edge dipole.}\label{sec:Qc}
The microscopic charge density $\rho(\vr)$ of a crystal can change rapidly on a
scale comparable or smaller than the size of its unit cell $a_i$. Hence, $\rho(\vr)$ itself
is not a physical observable but rather $\rho^\text{macro}(\vr)$,~\footnote{It is
easy to check that the first (second) moments of the charge densities
$\rho(\vr)$ and $\rho^\text{macro}(\vr)$ are the same if the first (first and
second) moments of $g(\vr)$ vanish.} obtained by spatial averaging
(convolution) from $\rho(\vr)$,~\cite{jackson1999,vanderbilt2018}
\begin{align}
	\rho^\text{macro}(\vr)=\int d^2 r^\prime \rho(\vr^\prime)g(\vr-\vr^\prime),
	\label{eq:rho_m}
\end{align}
where $g(\vr)$ is a normalized function, positive in the vicinity of $\vr=0$.
One possible choice is a Gaussian function, $g(\vec
r)=e^{-\frac{r^2}{2\xi^2}}/(\pi\xi^2)$, where the spread $\xi>a_i$ should be
chosen with some care. Namely, the crystal's charge neutrality implies that
$\rho^\text{macro}(\vr)$ vanishes for $\vec r$ away from the crystal boundaries,
which is satisfied with given accuracy only for a sufficiently large $\xi$.

As an example, consider a tight-binding model with eigenvectors
$\vert\psi_n\rangle$. The microscopic charge distribution $\rho(\vr)$ is
obtained from the projector onto occupied states (i.e., ground-state density
operator)
\begin{align}
	{\cal P}(\vx,\vx^\prime)&=\sum_{n\in\text{occ}}\psi_n^*(\vx)\psi_n(\vx^\prime),
	\label{eq:P}
\end{align}
as
\begin{align}
	\rho(\vr)&=-e\sum_\vx {\cal P}(\vx,\vx)\delta(\vr-\vx)+\rho^\text{ion}(\vr).
	\label{eq:rhor}
\end{align}
In the above equations, $\vx$ runs over the sites of the tight-binding model,
$\rho^\text{ion}(\vr)$ is ionic charge distribution, and we modelled 
the charge distribution of electronic orbitals by $\delta$ function. The above
microscopic charge density varies rapidly within the unit cell.
\begin{figure}[t]
	\centering
	\includegraphics[width=0.8\columnwidth]{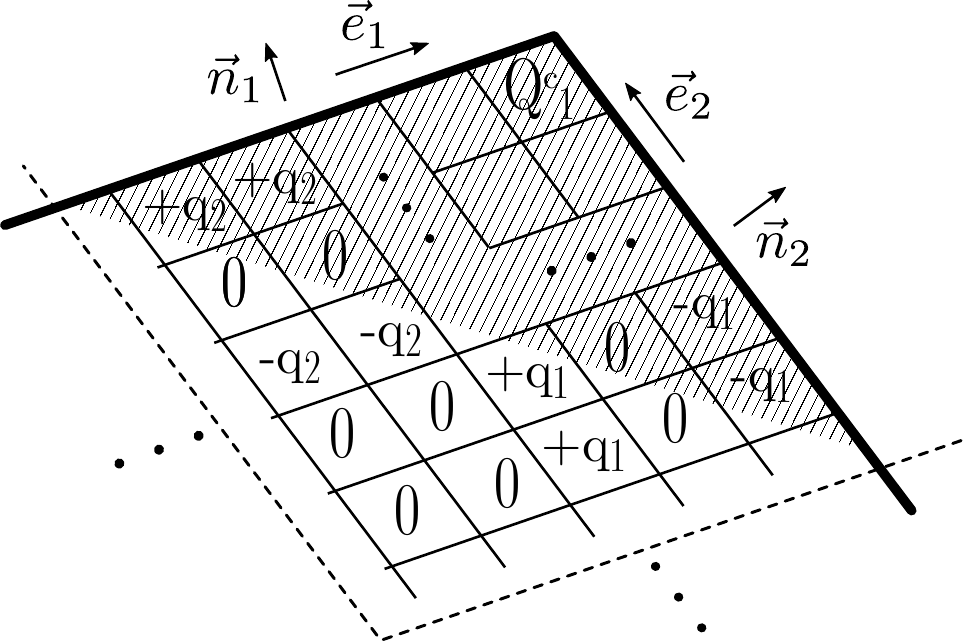}
	\caption{The macroscopic charge density $\rho^\text{macro}(\vr)$ for
	$\vr$ in the bulk, the two edges, and the corner regions. For clarity,
	the crystal is divided into cells (containing many unit cells) and the
	total charge in each cell is shown.  The corner is defined by the two
	edges with the unit normal vectors $\vec n_\alpha$, $\alpha=1,2$.  When
	either $D^\text{edge}_1$ or $D^\text{edge}_2$ are non-vanishing,
	integrating $\rho^\text{macro}$ over different corner regions ${\cal
	R}^\text{c}$, denoted by dashed lines and the hatched region, results
in different corner charge.}
	\label{fig:Qc}
\end{figure}

In practice, even the macroscopic charge density $\rho^\text{macro}(\vr)$ of a
crystal in not directly measured but rather the features that can be extracted
from it: the end or corner charges and the edge dipoles. Let us first consider
a finite quasi-one-dimensional system with charge density
$\rho^\text{macro}(r_1)$, $\vr=r_1\va_1$. Because of charge neutrality in the bulk,
$\rho^\text{macro}(r_1)$ is non-vanishing only for $r_1$ close to the ends of
the crystal. We define the edge charge $Q^\text{e}$
\begin{align}
	Q^\text{e}&=a_1\int dr_1 \rho^\text{macro}(r_1),
	\label{eq:Qe}
\end{align}
where the integration region includes only one end. For a tight-binding model,
$Q^\text{e}$ can be obtained from Eqs.~(\ref{eq:rhor}) and (\ref{eq:rho_m})
using Gaussian function with appropriately chosen $\xi$. Alternatively, as a
more straightforward approach, one performs moving window average on
$\rho$ which corresponds to the choice of $g(r_1)$ in Eq.~(\ref{eq:rho_m})
to be unit-box function, having value $1/a_1$ within the unit cell at origin and
zero otherwise.  The result of this averaging procedure is (see Sec.~4.5 of
Ref.~\onlinecite{vanderbilt2018})
\begin{align}
	Q^\text{e}&=a_1\int_{-\infty}^\infty dr_1 f_1(r_1)\rho(r_1),
	\label{eq:Qe_ramp}
\end{align}
where $f_\alpha(r)$ is the following ramp function,
\begin{align}
	f_\alpha(r)&=
	\begin{cases}
		0 &  r<r_{0\alpha}\\
		r-r_{0\alpha} &  r_{0\alpha}\le r\le r_{0\alpha}+1\\
		1 & \text{otherwise} ,
	\end{cases}
	\label{eq:falpha}
\end{align}
with $r_{0\alpha}$ far away from the ends.

Next, we consider finite two-dimensional (insulating) crystal that is charge
neutral and with vanishing edge charge density. Under these assumptions,
$\rho^\text{macro}(\vr)$ has to vanish for $\vr$ away from the boundaries, for
$\vr$ close to the middle of the edges there can be two spatially separated
line charge densities of opposite signs, while for $\vr$ close to the corners
the macroscopic charge density is generally nonvanishing; see
Fig.~\ref{fig:Qc}. Two features can be extracted from such
$\rho^\text{macro}(\vr)$: edge dipole and corner charge. The edge dipole
$D^\text{edge}_\alpha$, for the edge along the lattice vector $\ve_\alpha$ and with
the unit normal vector $\vec n_\alpha$, is defined as
\begin{align}
	D^\text{edge}_\alpha&=L_\alpha\int d(\vr\cdot\vec n_\alpha)\vr\cdot\vec n_\alpha \rho^\text{macro}(\vr),
	\label{eq:Dedge}
\end{align}
where the integration line crosses around the middle of the edge and
$L_\alpha=\vert\ve_\alpha\vert$. The corner charge $Q_\alpha^\text{c}$ is
defined as integral of macroscopic charge density over certain corner region
${\cal R}_\alpha^\text{c}$
\begin{align}
	Q_\alpha^\text{c}&=\int_{ {\cal R}_\alpha^\text{c}} d^2r\rho^\text{macro}(\vr).
	\label{eq:Qcdef}
\end{align}
Only for vanishing $D^\text{edge}_\alpha$ and $D^\text{edge}_{\alpha+1}$ is
$Q_\alpha^\text{c}$ independent of the choice of ${\cal R}_\alpha^\text{c}$;
see Fig.~\ref{fig:Qc}. The quantities~(\ref{eq:Dedge}) and (\ref{eq:Qcdef}) can be
computed directly from microscopic charge density $\rho(\vr)$ if a moving window
average is performed with the ``window'' corresponding to the unit cell defined
by the corresponding corner. The result for edge dipole is
\begin{align}
	D^\text{edge}_\alpha&=A_\text{cell}\int_{r_{0\alpha}}^{r_{0\alpha}+1}dr_\alpha\int dr_{\alpha+1}\vec r\cdot\vec n_\alpha\rho(\vr),
	\label{eq:Dedgemicro}
\end{align}
with $\vr=r_\alpha \ve_\alpha+r_{\alpha+1}\ve_{\alpha+1}$ and
$A_\text{cell}=\vert\ve_\alpha\times\ve_{\alpha+1}\vert$, where $\ve_\alpha$
and $\ve_{\alpha+1}$ are the edge lattice vectors and the integration limits
for $r_{\alpha+1}$ enclose the edge. For corner region ${\cal R}^\text{c}$
marked by the dashed line in Fig.~\ref{fig:Qc}, one finds
\begin{align}
	Q_\alpha^\text{c}&=\int d^2r f_\alpha(r_\alpha)f_{\alpha+1}(r_{\alpha+1})\rho,
	\label{eq:Qc_ramp}
\end{align}
where the integration is over the whole space. These features of the crystal's
charge density, $D^\text{edge}_\alpha$ and $Q^\text{c}_\alpha$, can be
reproduced by representing the crystal as a collection of regions with
polarization $\vec P^\text{edge}_\alpha$ as in Fig.~\ref{fig:1}. These edge
polarizations $\vec P^\text{edge}_\alpha$ are not uniquely determined by
the crystal's charge density---while their transversal component is given by the
corresponding edge dipole
\begin{align}
	\vec P^\text{edge}_\alpha\cdot\vec n_\alpha&=D^\text{edge}_\alpha,
	\label{eq:PedgeDedge}
\end{align}
it is only the sum of the longitudinal components that is fixed by the corner
charge via bulk-and-edge to corner correspondence~(\ref{eq:bcc}). In
Sec.~\ref{sec:main} we present an algorithm for computing the edge polarizations
$\vec P^\text{edge}_\alpha$ from the knowledge of the crystal's bulk and
the corresponding ribbon band structures.

\subsection{Modern theory of electric polarization. Bulk-boundary correspondence.}\label{sec:modth}
\begin{figure}[t]
	\centering
	\includegraphics[width=0.9\columnwidth]{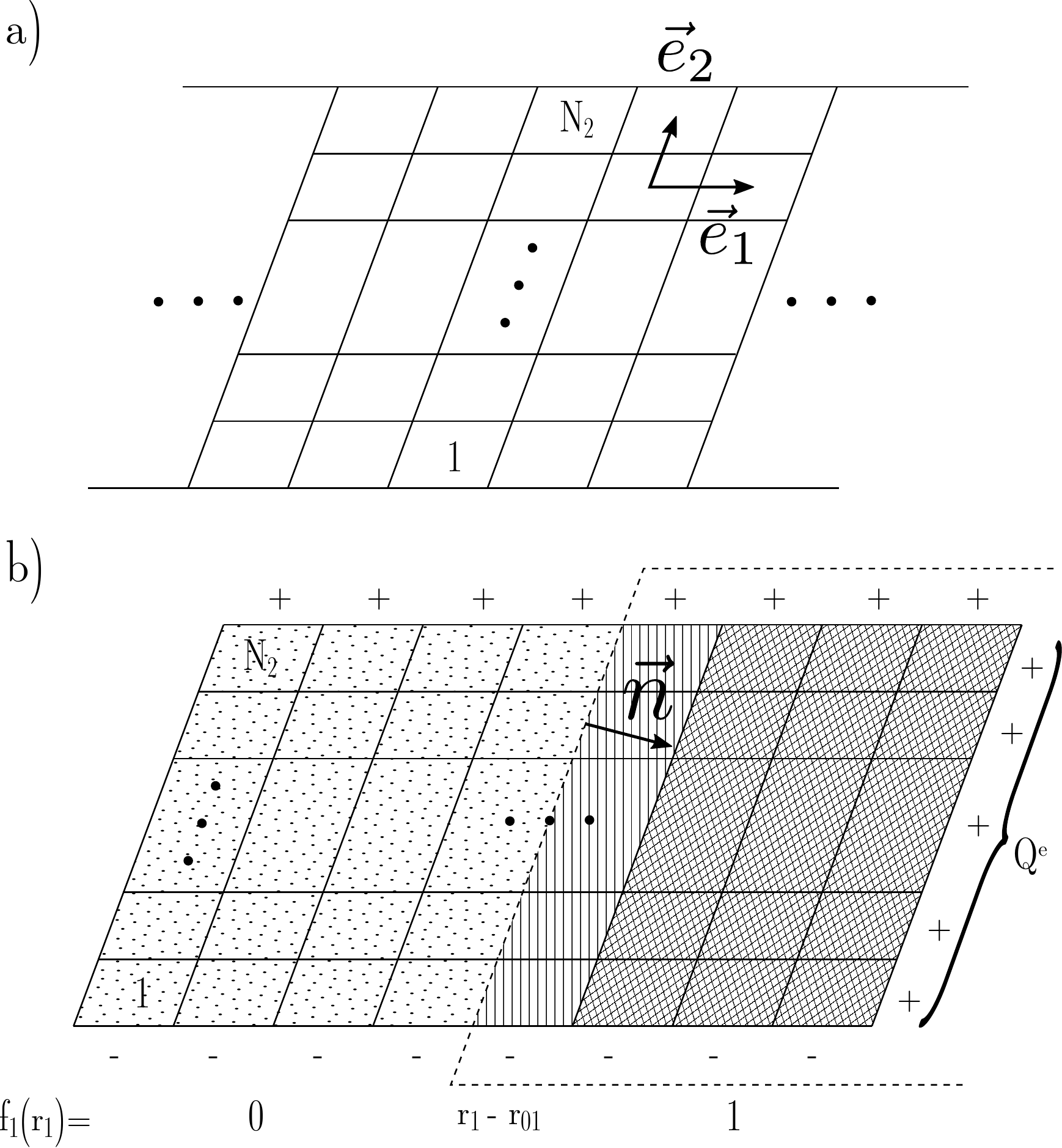}
	\caption{A ribbon infinite in $\vec e_1$ direction with $N_2$
		unit cells in $\vec e_2$-direction (a). Nonzero polarization
		$\vec P$ implies end charge $Q^\text{e}$ once the ribbon is
		terminated (b). The end charge $Q^\text{e}$, defined as the
		charge contained in the region bounded by the dashed lines, can
		be computed by multiplying the charge density $\rho$ in
		Eq.~(\ref{eq:Qe_ramp}) by continuous ramp function $f_1(r_1)$, which
		takes values $0$ (dotted region), $r_1-r_{01}$ (hatched region)
		and $1$ (filled region) (b), followed by integration over the
		whole space. The region where the ramp function takes non-zero
		values is denoted by dashed line $(\vec r-\vr_0)\cdot\vec n=0$,
		which crosses the ribbon far from ends.}
	\label{fig:2}
\end{figure}
For purposes of this work, we will be interested in polarization of a ribbon.
Consider a ribbon infinite in the $\vec e_1$ direction, with $N_2$ unit cells in
the $\vec e_2$ direction, where $\vec e_{1,2}$ are lattice vectors; see
Fig.~\ref{fig:2}a. We assume that the ribbon is described by a gapped
$(NN_2)\times(NN_2)$ Bloch Hamiltonian $h_{k_1}$, where $N$ is the number of
sites per unit cell. For each $k_1$ point, we denote the projector onto occupied
Bloch wavefunctions $\vert\psi_{nk_1}\rangle$ by ${\cal P}_{k_1}=\sum_{n=1}^{N_2
N_\text{occ}}\vert\psi_{nk_1}\rangle\langle\psi_{nk_1}\vert$, where the integer
$N_\text{occ}$ is the number of the occupied states per unit cell. For the
definition of ${\cal P}_{k_1}$, the scalar product is assumed to be taken
over the supercell only, i.e., ${\cal P}_{k_1}$ is $(NN_2)\times(NN_2)$ matrix
and ${\cal P}_{k_1+2\pi}={\cal P}_{k_1}$. Modern theory of electric
polarization states that the polarization $\vec P$ of the ribbon is given
by~\cite{king-smith1993,rhim2017}
\begin{align}
	\vec P=&\frac{e}{2\pi}\left( i\ln{\det}^\prime\left[ \prod_{k_1}{\cal P}_{k_1} \right]\vec e_1-\int_0^{2\pi}dk_1\text{Tr}\left[ \vec{\hat x} {\cal P}_{k_1} \right]\right),
	\label{eq:3}
\end{align}
where ${\det}^\prime U$ denotes the product of non zero eigenvalues of $U$ and
$\vec{\hat x}$ is the position operator. The polarization~(\ref{eq:3}) depends
on the choice of the origin $\vr=0$, which can be avoided if the above
expression is modified to include the ionic contribution to the charge density.
The bulk-boundary correspondence states
\begin{align}
	Q^\text{e}\mod e&=L\vec P\cdot \vec n/\vert\vec e\times\vec e_1\vert,
	\label{eq:4}
\end{align}
where $\vec n$ is a unit vector, perpendicular to some lattice vector $\vec e$,
and $L=\vert\vec e\vert$ is the shortest repeated length along the direction
$\vec e$. The end charge $Q^\text{e}$, which is the macroscopic charge
contained in the dashed region in Fig.~\ref{fig:2}b, can be computed from
Eq.~(\ref{eq:Qe_ramp}).

An alternative formulation of electric polarization is obtained by expressing
the projector~(\ref{eq:P}) onto the occupied states of the ribbon as
\begin{align}
	{\cal P}=\sum_{R_1}{\cal P}_{R_1}=\sum_{R_1,n}\vert w_{R_1,n}\rangle\langle w_{R_1,n}\vert,
	\label{eq:WF}
\end{align}
where $\vert w_{R_1,n}\rangle$ are (nonunique) exponentially localized Wannier
functions (WFs) and $R_1$ enumerates different supercells of the ribbon. When
the ribbon is infinite (or under periodic boundary conditions), the shape of
WFs is independent of $R_1$ due to translational symmetry. The
polarization~(\ref{eq:3}) can be expressed using WFs as
\begin{align}
	\vec P&=-e\mathrm{Tr}\left[ {\cal P}_{R_1}\vec{\hat x} \right]\mod e\vec e_1,
	\label{eq:PWF}
\end{align}
which is independent of $R_1$.

\section{Bulk-and-edge to corner correspondence}\label{sec:main}
\begin{figure}[b]
	\centering
	\includegraphics[width=\columnwidth]{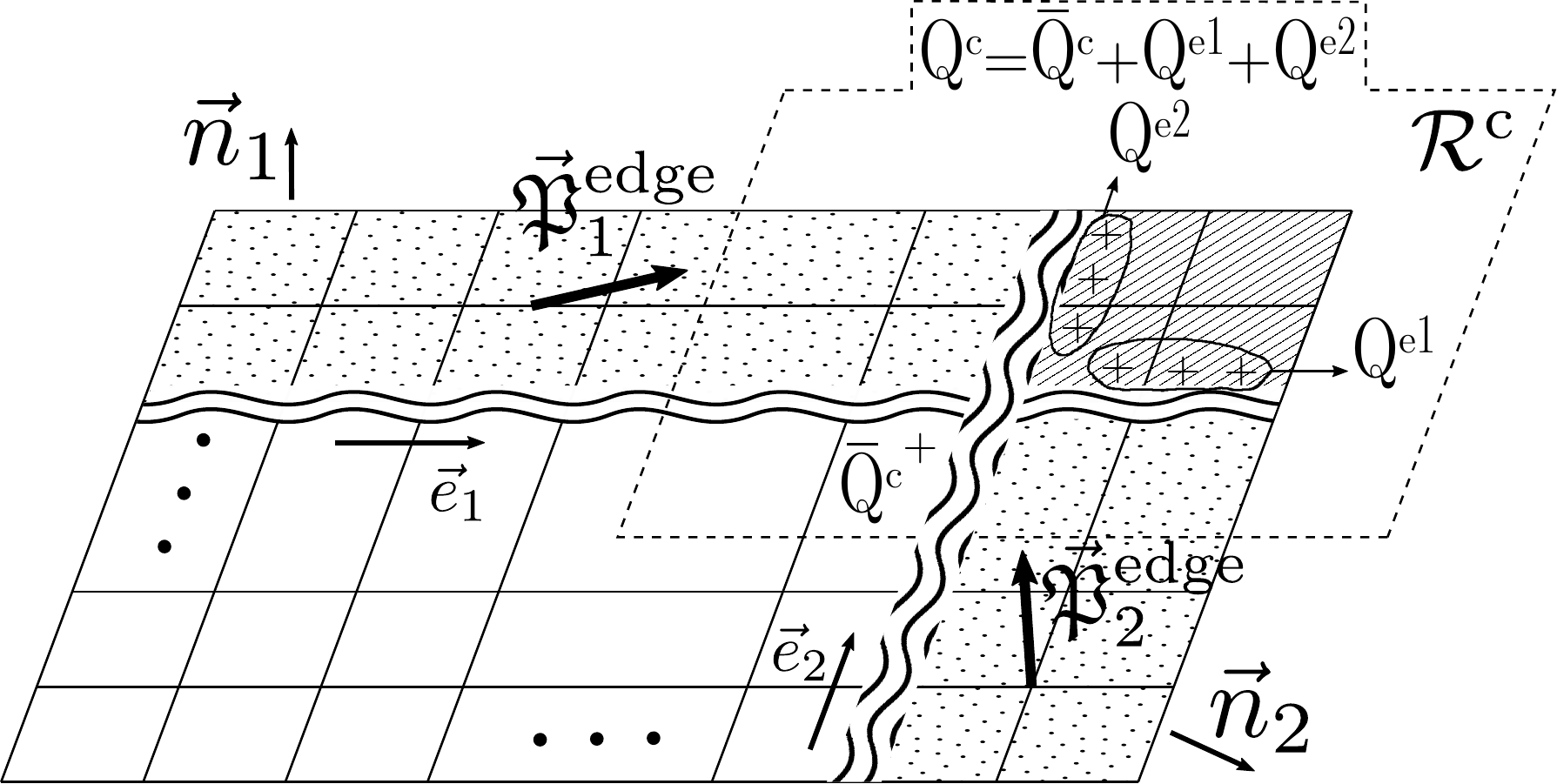}
	\caption{A charge neutral flake with total bulk polarization zero. The flake
		is semi-infinite in both directions. Two cuts (wavy lines)
		along a charge neutral lines $(\vr-\vr_0)\cdot\vec n_\alpha=0$,
		$\alpha=1,2$, split the flake into four
		subsystems: the two edges (dotted regions), the corner (hatched
		region) and the remaining bulk subsystem.  A fractional part
		of the upper-right corner charge is given by Eq.~(\ref{eq:6}). If
		the cuts are chosen to be Wannier cuts, the resulting
		polarization of the edge subsystem is called Wanner edge
		polarization $\vP^\text{edge}_\alpha$. The corner region ${\cal
		R}^\text{c}$ is enclosed by dashed line. The end charge of the
		two edge subsystems is denoted by $Q^\text{e1}$ and
		$Q^\text{e2}$.}
		\label{fig:3}
\end{figure}
In this section, we prove the correspondence~(\ref{eq:bcc}) between the
corner charge and the bulk-and-edge. We consider a terminated system (flake)
with vanishing bulk polarization and focus on the upper-right corner, where for
the purpose of the following discussion we may assume that the remaining three
corners lie at infinity. We use the notation where two edges have unit normal
vectors $\vec n_\alpha$ and are along certain lattice vectors $\vec e_\alpha$,
$\alpha=1,2$. The lattice vectors $\ve_1$ and $\ve_2$ define the unit-cell area
$A_\text{cell}=\vert\vec e_1\times\vec e_2\vert$; see Fig.~\ref{fig:3}. The
reduced coordinates $(r_1,r_2)$ are defined as $\vr=r_1\ve_1+r_2\ve_2$.

We make two cuts along the lines $(\vr-\vr_0)\cdot\vec
n_\alpha=0$, where the point $\vr_0$ lies in the bulk.  These cuts divide the
flake into four regions (subsystems) (see Fig.~\ref{fig:3}): the two
half-infinite edges (dotted regions), the corner (hatched region), and the
remaining bulk region.  Denoting the corner charge of the bulk subsystem by
$\bar Q^\text{c}$, we express the flake's corner charge $Q^\text{c}$ as
\begin{align}
	Q^\text{c}&=\bar Q^\text{c}+Q^{\text{e}1}+Q^{\text{e}2}\mod e,
	\label{eq:6}
\end{align}
where $Q^{\text{e}1,2}$ are the end charges of the two edge subsystems; see
Fig.~\ref{fig:3}.  In writing above relation, we used that the corner-subsystem
is charge neutral. (More generally, it contains integer multiple of electron
charge $e$.)

While many cuts allow one to separate flake into four subsystems, we
additionally require that $\bar Q^\text{c}$ is a bulk quantity and
$\vP^\text{edge}_\alpha$ are edge quantities. To be more precise, we require
that $\bar Q^\text{c}$ is computable in terms of the bulk band structure, and
similarly $\vP^\text{edge}_\alpha$ should be computable from the ribbon band
structure for the ribbon along $\vec e_\alpha$. We call such cuts 
\textit{Wannier cuts}, and the resulting edge polarizations
$\vP^\text{edge}_\alpha$ \textit{Wannier edge polarizations}. For a Wannier
cut, the relation~(\ref{eq:6}) becomes bulk-and-edge to corner correspondence,
\begin{align}
	Q^\text{c}&=\bar Q^\text{c}+\frac{L_2}{A_\text{cell}}\vP_1^\text{edge}\cdot\vec n_2+\frac{L_1}{A_\text{cell}} \vP_2^\text{edge}\cdot\vec n_1\mod e.
\end{align}

Below we first define the bulk subsystem and prove that the resulting $\bar
Q^\text{c}$ is expressed in terms of quadrupole tensor of the charge density of
the Wannier functions. Subsection~\ref{sec:Pedge} details on how to compute
\textit{Wannier edge polarization} $\vP^\text{edge}_\alpha$.

\subsection{Bulk-subsystem.}\label{sec:Wcutter}
To define the bulk subsystem, we make a choice for bulk WFs, $\bar w_{\vec R
n}(\vec x)$. Assuming that WFs are assigned to their home unit cell, the charge
density of the bulk WFs $\rho^\text{WF}(\vr)$, centered around $\vr=0$, reads
\begin{align}
	\rho^\text{WF}(\vec r-\vec R)&=-e\sum_{\vx n}\vert\bar w_{\vec R n}(\vec x)\vert^2\delta(\vec r-\vec x)+\rho_{\vec R}^\text{ion}(\vec r),
	\label{eq:rhoWF}
\end{align}
where we included the ionic contribution $\rho^\text{ion}_{\vec R}$ to the
unit cell at $\vec R$.  The second moment of the above charge density defines
the (bulk) quadrupole tensor density $\hat q$,
\begin{align}
	\hat q&=\frac{1}{A_\text{cell}}\sum_{\alpha,\beta=1,2}\hat q_{\alpha\beta}\vec e_\alpha\otimes\vec e_\beta,
	\label{eq:hatq}
\end{align}
where ``$\otimes$'' denotes tensor product and $\hat q_{\alpha\beta}=\langle
r_\alpha r_\beta\rangle_{\rho^\text{WF}}\equiv\int d^2r\rho^\text{WF}(\vr)
r_\alpha r_\beta$.

\begin{figure}[h]
	\centering
	\includegraphics[width=.8\columnwidth]{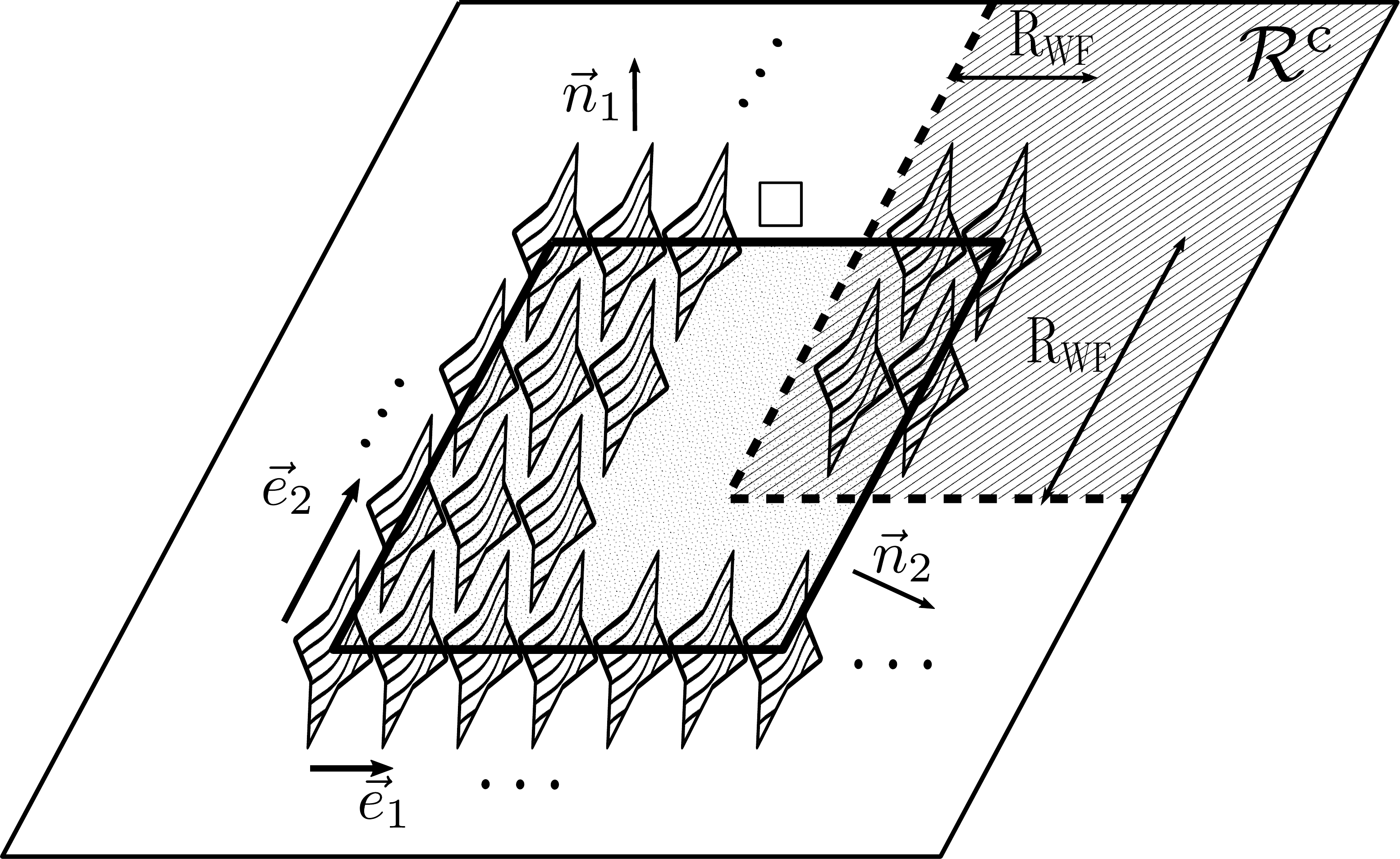}
	\caption{A flake with the regions $\square$ (dotted region with a thick
		solid line) and ${\cal R}^\text{c}$ (hatched region with dashed
		line). The bulk subsystem, located away from the edges (thin
		solid lines), is obtained by tiling $\square$ with the occupied
		bulk WFs (wavy star-shaped objects). The integer $R_\text{WF}$
		is the ``radius'' of the bulk WFs, outside of it the charge density
	of the bulk WFs can be neglected.}
	\label{fig:4}
\end{figure}

We now go back to the flake from Fig.~\ref{fig:3}, and select the rectangle
$\square$ in the reduced coordinates $(r_1,r_2)$. We define the bulk subsystem
to consist of the flake's WFs with the center in the rectangle $\square$; see
Fig.~\ref{fig:4}. For the rectangle $\square$ deep in the bulk, the
bulk-subsystem is obtained by tiling the rectangle $\square$ with the
\textit{bulk} WFs.  Hence, its charge density $\bar\rho(\vec x)$ takes a simple
form:
\begin{align}
	\bar\rho(\vec r)=&\sum_{\vec R\in\square}\rho^\text{WF}(\vec r-\vec R).
	\label{eq:10}
\end{align}
Note that the charge density $\bar\rho$ extends beyond the rectangle $\square$.

We now prove that the corner charge and the edge dipole of the bulk subsystem,
$\bar Q^\text{c}_\alpha$ and $\bar D^\text{edge}_\alpha$, are expressed in
terms of the bulk quadrupole tensor $\hat q$. For simplicity, we assume that all
sites of the flake lie on the lattice itself (lattice without the basis). In
this case, instead of working with $\rho^\text{WF}(\vr)$, we use the quantity
$Q^\text{WF}_{\vec R}$: the charge at lattice position $\vec R$, originating
from the bulk Wannier functions centered at $\vec R=0$,
\begin{align}
	Q^\text{WF}_{\vec R}&=-e\sum_{n=1}^{N_\text{occ}}\vert w_{(0,0)n}(\vec R)\vert^2+N_\text{occ}e\delta_{R_1,0}\delta_{R_2,0}.
	\label{eq:12p}
\end{align}
In the above expression, we assume that the ionic charge contribution $e
N_\text{occ}$ is localized at the lattice sites, and thus $\sum_\vR
Q^\text{WF}_\vR=0$. We compute $\bar Q^\text{c}$ using the corner region ${\cal
R}^\text{c}$ with boundaries $(\vr-\vr_0)\cdot\vec n_\alpha=0$, where $\vr_0$
lies in the bulk; see Fig.~\ref{fig:4}. Hence, the corner charge of the
bulk subsystem $\bar Q^\text{c}$ is obtained from Eq.~(\ref{eq:Qc_ramp}),
\begin{align}
	\bar Q^\text{c}=\sum_{\vec R\in\square}&\left[ -\theta(-R_1)R_{\text{WF}}-\theta(-R_2)R_{\text{WF}}+\theta(-R_1)\theta(-R_2)\right.\nonumber\\
	&+\theta(R_1)(R_{\text{WF}}-\theta(-R_2))+\theta(R_2)(R_{\text{WF}}-\theta(-R_1))\nonumber\\
&\left.+\theta(R_1)\theta(R_2)\right] Q^\text{WF}_{\vec R},
	\label{eq:12}
\end{align}
where $\theta(x)\equiv\vert x\vert H(x)$ with $H(x)$ being the Heaviside step
function and $R_{\text{WF}}$ being the radius of the bulk WFs, see
Fig.~\ref{fig:4}.  Since $\sum_\vR Q^\text{WF}_\vR=0$ holds, only the bulk WFs
whose charge is not fully contained in ${\cal R}^\text{c}$ contribute to $\bar
Q^\text{c}$.  The three terms in the first line of the sum~(\ref{eq:12}) count
the charge contained in the region outside of ${\cal R}^\text{c}$ (not
necessarily inside of $\square$), originating from the bulk WFs whose center is
in ${\cal R}^\text{c}$.  Similarly, the second and the third lines of the
sum~(\ref{eq:12}) correspond to the charge contained within ${\cal
R}^\text{c}$, originating from the bulk WFs whose center is in
$\square\setminus{\cal R}^\text{c}$. Using the assumption that the bulk
polarization vanishes $\vec P=\sum_{\vec R}\vec R Q^\text{WF}_{\vec R}=0$, we
rewrite Eq~(\ref{eq:12}) as
\begin{align}
	\bar Q^\text{c}&=\sum_{\vec R\in\square}R_1R_2Q^\text{WF}_{\vec R}=\frac{L_1L_2}{A_\text{cell}}\vec n_1\cdot\hat q\cdot\vec n_2,
	\label{eq:13}
\end{align}
where in the last equality we used Eq.~(\ref{eq:hatq}), $L_\alpha=\vert\vec
e_\alpha\vert$, and that the unit normal vectors $\vec n_\alpha$ point toward
the outside of the subsystem. The above result is valid for both convex and concave
angles $\angle(\vec e_1,\vec e_2)$.

To obtain the edge dipole $\bar D^\text{edge}_1$ from the charge density
$\bar\rho$, we consider a flake infinite in the $\vec e_1$ direction (ribbon)
and focus on the upper edge. For concreteness we assume that the unit cells at
the top edge of $\square$ have coordinates $(R_1,-1)$. Because of
translational invariance, the charge $\mathfrak{Q}_{R_2}$, which is
(microscopic) charge of the ribbon at the lattice site $\vec R$, is
independent of $R_1$.  Using Eqs.~(\ref{eq:10}) and (\ref{eq:12p}) we write,
\begin{align}
	\mathfrak{Q}_{R_2}&=
	\begin{cases}
		\sum_{R_2^\prime>R_2}Q^\text{WF}_{R_2^\prime} & R_2\ge0,\\
		-\sum_{R_2^\prime\le R_2}Q^{\text{WF}}_{R_2^\prime} & R_2<0,
	\end{cases}
	\label{eq:frakQ}
\end{align}
where the notation $Q^\text{WF}_{R_2}\equiv\sum_{R_1}Q^\text{WF}_{\vec R}$ has
been introduced. The edge dipole [see Eq.~(\ref{eq:Dedgemicro})] for the edge along $\vec
e_1$ is expressed as
\begin{align}
	\frac{\bar D^\text{edge}_1}{\ve_2\cdot\vec n_1}&=\sum_{R_2}R_2\mathfrak{Q}_{R_2}\nonumber\\
	&=-\sum_{R_2\ge0}R_2\sum_{-R_2<R_2^\prime\le R_2}Q^\text{WF}_{R_2^\prime}.
	\label{eq:Dedgesquare}
\end{align}
Changing the order of the two sums in the above expression, we obtain
\begin{align}
	\frac{\bar D^\text{edge}_1}{\ve_2\cdot\vec n_1}=&\sum_{R_2^\prime\ge0}Q^\text{WF}_{R_2^\prime}\sum_{R_2^\prime\le R_2<R_\text{WF}}R_2-\sum_{R_2^\prime<0}Q^\text{WF}_{R_2^\prime}\sum_{R_2^\prime\le R_2<R_\text{WF}}R_2\nonumber\\
	&=\frac{1}{2}\sum_{R_2^\prime}{R_2^\prime}^2 Q^\text{WF}_{R_2^\prime},
	\label{eq:intermediate}
\end{align}
where the cutoff $R_\text{WF}$ drops out for the ribbon wider than the radius
of the bulk WFs, and in writing the second line we used
$\sum_{R_2^\prime}Q^\text{WF}_{R_2^\prime}=\sum_{R_2^\prime}R_2^\prime
Q^\text{WF}_{R_2^\prime}=0$. Repeating the same calculation for the other
edge, we obtain
\begin{align}
	\bar D^\text{edge}_\alpha=L_\alpha\vec n_\alpha\cdot\hat q\cdot\vec n_\alpha/2.
	\label{eq:DedgeB}
\end{align}
\begin{figure}[h]
	\centering
	\includegraphics[width=.5\columnwidth]{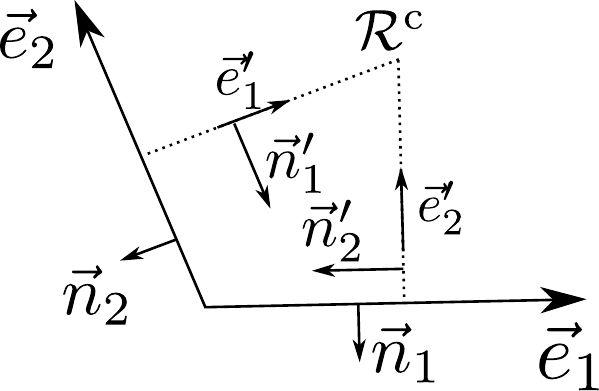}
	\caption{A corner defined by the edges along lattice vectors
		$\ve_{1,2}$ and unit-normal vectors $\vec n_{1,2}$. The corner
		region ${\cal R}^\text{c}$ is defined by the (dashed) lines
	$(\vr-\vr_0)\cdot\vec n_\alpha^\prime=0$, where the point $\vr_0$ lies
in the bulk.}
	\label{fig:Rc}
\end{figure}

The relations~(\ref{eq:13}) and (\ref{eq:DedgeB}) show that the corner charge
and the edge dipole resulting from the macroscopic average of the charge
density $\bar\rho$, can be seen to be be generated by polarizations
$L_\alpha\hat q\cdot\vec n_\alpha/2$, placed long the four edges of $\square$.
For a more general case, when the corner region ${\cal R}^\text{c}$ is defined
by lines $(\vr-\vr_0)\cdot\vec n_\alpha^\prime=0$, applying the bulk-boundary
correspondence~(\ref{eq:4}) to these edge polarizations gives
\begin{align}
	\bar Q^\text{c}&=\frac{L_1\vert \vec e_2^\prime\vert}{\vert\vec e_1\times\vec e_2^\prime\vert}\vec n_2^\prime\cdot\hat q\cdot\vec n_1/2+\frac{\vert\vec e_1^\prime\vert L_2}{\vert\vec e_1^\prime\times\vec e_2\vert}\vec n_1^\prime\cdot\hat q\cdot\vec
	n_2/2,
	\label{eq:Qcgen}
\end{align}
where $\vec e_\alpha^\prime$ are lattice vectors satisfying $\vec
e_\alpha^\prime\cdot\vec n_\alpha^\prime=0$, normalized such that $\vert\vec
e_\alpha^\prime\vert$ gives the shortest repeated length in the lattice direction
$\vec e_\alpha^\prime$, see Fig.~\ref{fig:Rc}. The above expression reduces to
Eq.~(\ref{eq:13}) for $\vec n_\alpha^\prime=\vec n_\alpha$.

\subsection{Edge-subsystem. Wannier edge polarization \texorpdfstring{$\vP_\alpha^\text{edge}$}{TEXT}.}\label{sec:Pedge}
To define the edge subsystem and Wannier edge polarization, we considering a
ribbon with the periodic boundary condition in the $\vec e_\alpha$ direction. We
call ``Wannier cut'' the procedure where the bulk Wannier functions, used to
define the bulk subsystem in the previous subsection, are removed from the
middle of the ribbon. After such removal, what remains of the ribbon are two
edge subsystems. For a sufficiently wide cut, the charge densities of the two
edge subsystems do not overlap when projected onto
the $\ve_{\bar\alpha}$ direction. Therefore, we can define the subsystem
corresponding to a single edge $\alpha$ and its polarization we call Wannier
edge polarization $\vP_\alpha^\text{edge}$.

For concreteness, we set $\alpha=1$ and denote by $h_{k_1}$ the Bloch
Hamiltonian of the ribbon, with supercell having $2N_2+1$ unit cells located at
positions $R_2\vec e_2$, $R_2\in[-N_2,N_2]$.  The translationally invariant Wannier
cut is performed using hybrid bulk WFs $\vert \bar w_{k_1R_2n}\rangle$,
\begin{align}
	\vert\bar w_{k_1R_2n}\rangle&=\sum_{R_1=-N_1}^{N_1-1}e^{ik_1R_1}\vert\bar w_{\vec Rn}\rangle,	
	\label{eq:15}
\end{align}
where the periodic boundary condition identifies the sites at $R_1=-N_1$ with those
at $R_1=N_1$.  The Wannier cut is performed in the middle of the ribbon
supercell by removing $2L+1$ hybrid bulk WFs $\vert\bar w_{k_1R_2n}\rangle$
with $R_2\in[-L,L]$ from the space spanned by the occupied states
\begin{align}
	{\cal P}_{k_1}^L&\equiv {\cal P}_{k_1}-\sum_{\substack{n=1\\R_2=-L}}^{\substack{n=N_\text{occ}\\R_2=L}}\vert\bar w_{k_1R_2n}\rangle\langle\bar w_{k_1R_2n}\vert.
	\label{eq:17}
\end{align}
The integer $L$ should be chosen sufficiently large such the projector
$P_{k_1}^L$ does not contain the sites from the middle unit cell of the
supercell, i.e., 
\begin{align}
	{\cal P}_{k_1}(\vec x,\vec x^\prime)\rightarrow 0,
	\label{eq:middle_layer0}
\end{align}
for the reduced coordinates $x_2,x_2^\prime\in [-1/2,1/2]$. As WFs are more
localized, a smaller value of $L$ is required. The matrix elements of the
projector onto occupied states of the edge subsystem above the Wannier cut read
\begin{align}
	{\cal P}^\text{edge}_{k_1}(\vec x,\vec x^\prime)&=H(x_2)H(x_2^\prime){\cal P}_{k_1}^L(\vec x,\vec x^\prime).
	\label{eq:cut0}
\end{align}
Note that for a ribbon with the fixed width $N_2$, the value of $L$ should not
be too large; otherwise the hybrid bulk WFs do not fully belong to the space of
occupied states of the ribbon. This can be diagnosed by inspecting the charge
neutrality of the resulting edge subsystem's supercell,
\begin{align}
	\sum_{\vec x}\int\frac{dk_1}{2\pi}{\cal P}^\text{edge}_{k_1}(\vec x,\vec x)\rightarrow N_\text{occ}(N_2-L),
	\label{eq:neutrality}
\end{align}
where the value on the right-hand side is the ionic charge---the removal of the
hybrid bulk WF also removes the corresponding ionic charge. If the
conditions~(\ref{eq:middle_layer0}) and~(\ref{eq:neutrality}) are satisfied
with required accuracy, the Wannier cut has been performed successfully.
The edge polarization $\vP^\text{edge}_1$ is given by Eq.~(\ref{eq:3}) using
${\cal P}^\text{edge}_{k_1}$ in place of ${\cal P}_{k_1}$.

\subsection{Discussion}\label{ssec:discussion}
Putting together the bulk and the edge subsystems, we obtain the main
result of our work, namely that the corner charges and the edge dipoles are
determined by the edge polarizations~(\ref{eq:Pedge}). The bulk-and-edge to
corner charge correspondence is obtained after substituting Eq.~(\ref{eq:Qcgen})
and the expression for the edge polarization into Eq.~(\ref{eq:6}),
\begin{align}
	Q^\text{c}_{ {\cal R}^\text{c}}&=\frac{\vert \vec e_2^\prime\vert}{\vert\vec e_2^\prime\times\vec e_1\vert} \vec P^{\text{edge}}_1\cdot\vec n_2^\prime+\frac{\vert\vec e_1^\prime\vert}{\vert\vec e_1^\prime\times\vec e_2\vert}\vec P^{\text{edge}}_2\cdot\vec n_1^\prime,
	\label{eq:qcvb}
\end{align}
where the corner charge is defined by the corner region ${\cal R}^\text{c}$ in
Fig.~\ref{fig:Rc}.  It is worth mentioning that not only Wannier polarization
$\vP^\text{edge}_\alpha$ but also the edge polarization~(\ref{eq:Pedge})
depends on the choice of the bulk WFs. This observation agrees with the previously
mentioned statement that the two edge polarizations $\vec P^\text{edge}_\alpha$
have four independent components whereas there are only three independent
physical observables.

The bulk subsystem defined in Sec.~\ref{sec:Wcutter} is special because its
charge density can be obtained by tiling. The bulk subsystem can be viewed as a
flake with a special termination that we call \textit{Wannier termination}. It
is instructive to compare the relation between the corner charge and the
quadrupole moment of a flake with an arbitrary termination versus the one with
Wannier termination. Assuming that the two lowest moments of the flake's
(microscopic) charge density $\rho^\text{flake}$ vanish, we write the second
(off-diagonal) moment as
\begin{align}
	\hat q^\text{flake}_{12}&=\frac{1}{N_1N_2A_\text{cell}}\langle x_1x_2\rangle_{\rho^\text{flake}},
	\label{eq:qflake}
\end{align}
where the flake has $N_{1,2}$ unit cells along the two primitive vectors. For
an inversion-symmetric rhomboid flake, the macroscopic charge density
$\rho^{\text{flake},\text{macro}}$ consists of four corner charges $Q^\text{c}$
with alternating signs superimposed with the edge dipoles, see
Sec.~\ref{sec:Qc}. Denoting the coordinates of the center of charge of the
top-right corner by $\vec X=\frac{1}{2}X_1\vec e_1+\frac{1}{2}X_2\vec e_2$, with
the origin at the flake's inversion center, gives $\langle
x_1x_2\rangle_{\rho^{\text{flake},\text{macro}}}=\langle
x_1x_2\rangle_{\rho^{\text{flake}}}=Q^\text{c}X_1X_2$. The corner charge
$Q^\text{c}$ can be approximated from the flake's quadrupole moment,
\begin{align}
	Q^\text{c}&\sim \frac{X_1X_2}{N_1N_2}Q^\text{c}=\frac{L_1L_2}{A_\text{cell}}\vec n_1\cdot\hat q^\text{flake}\cdot\vec n_2.
	\label{eq:Qcapprox}
\end{align}
We used that $X_\alpha/N_\alpha \rightarrow 1$ in the thermodynamic limit
$N_\alpha\rightarrow\infty$. Hence, for a flake with an arbitrary termination,
the corner charge can be obtained from the microscopic charge density
$\rho^\text{flake}$ from Eqs.~(\ref{eq:qflake}) and~(\ref{eq:Qcapprox}) only
with algebraic accuracy in the flake's size. On the other hand, for a flake
with Wannier termination (i.e., bulk subsystem), $\hat q^\text{flake}$ is given
by Eq.~(\ref{eq:hatq}), and we know that the relation~(\ref{eq:13}) holds with
exponential accuracy as the flake size is increased beyond the radius of the
tile (Wannier functions). In other words, Wannier termination pins $X_1$
($X_2$) to the value $N_1$ ($N_2$).

\section{Examples}\label{sec:examples}
In this section, we consider three two-dimensional tight-binding models which we
use to illustrate the procedure described in the previous section. For each
example, we perform two independent calculations: (1) After diagonalization of
the flake's Hamiltonian, we obtain macroscopic charge density
$\rho^\text{macro}$ from Eqs.~(\ref{eq:rho_m}) and~(\ref{eq:rhor}), and
subsequently the corner charge $Q^\text{c}$ and the edge dipole $D^\text{edge}$
from Eqs.~(\ref{eq:Qcdef}) and~(\ref{eq:Dedge}), and (2) we make a choice of the
occupied bulk WF, calculate the bulk quadrupole tensor and the Wannier edge
polarizations for each edge of interest---these quantities give the edge
polarization~(\ref{eq:Pedge}) that are used to compute the corner charge and
the edge dipole moments; see Eqs.~(\ref{eq:bcc}) and (\ref{eq:PedgeDedge}).
Alternatively, $D^\text{edge}_\alpha$ can be obtained from the corresponding
ribbon calculation, see Eq.~(\ref{eq:Dedgemicro}). The first example considers
the Benalcazar-Bernevig-Hughes (BBH) model~\cite{benalcazar2017,benalcazar2019}
with broken fourfold rotation symmetry such that the corner charge is no longer
quantized. In the dimerized limit of BBH model, the procedure from
Sec.~\ref{sec:main} is carried out analytically. The second example considers a
model with a single occupied band where the corners are formed by several
different orientations of the edges. The final example is meant to illustrate a
scenario where the bulk contribution to the edge polarization~(\ref{eq:Pedge})
is small, which explains why it was overlooked in Ref.~\onlinecite{zhou2015}.

To carry out the above-mentioned calculations, we need to specify the bulk
Hamiltonian and the edge boundary conditions. We consider a special form of
boundary conditions that we call ``theorist'' boundary conditions: The
Hamiltonian around the boundaries is assumed to be the same as the bulk
Hamiltonian with the hoppings to the missing sites set to zero. Theorist
boundary conditions are used here out of convenience; they do not have any
particular relevance for realistic systems.

\subsection{BBH model with broken fourfold rotation symmetry}
Here we consider a two-dimensional $\pi$-flux dimerized
model,~\cite{benalcazar2017,benalcazar2019} which is a particular case of BBH
model that exhibits quantized corner charge protected by fourfold rotation
symmetry ${\cal C}_4$
\begin{align}
	H=&e^{i\pi/4}\sum_{\vR}\left(\ket{\vR,1}\bra{\vR+\va_x,2}+\ket{\vR+\va_y,4}\bra{\vR,1}\right.\nonumber\\
	&+\ket{\vR+\va_x+\va_y,3}\bra{\vR+\va_y,4}\nonumber\\
	&\left.+\ket{\vR+\va_x,2}\bra{\vR+\va_x+\va_y,3}\right)+\text{h.c.}+H_4,
	\label{eq:BBH}
\end{align}
where the four orbitals $\ket{\vR\gamma}$, $\gamma=1,\dots,4$ are placed on a
square lattice with primitive vectors $\va_\alpha=a\ve_\alpha$, where $\ve_{x}$
($\ve_y$) is the unit vector along the $x$ ($y$) direction. The fourfold rotation
acts as ${\cal C}_4:\ket{\vR\gamma}\rightarrow \ket{\vR\gamma+1}$ and is
responsible for the quantization~\cite{benalcazar2019} of the corner charge to
the value of $Q^\text{c}=e/2\mod e$. The term $H_4$ breaks fourfold rotation
symmetry while preserving its twofold rotation subgroup
\begin{align}
	H_4=\delta\sum_\vR (-1)^\gamma \ket{\vR\gamma}\bra{\vR\gamma}.
	\label{eq:H4}
\end{align}
The two occupied bulk Wannier functions localized on four sites can be
written as
\begin{align}
	\bar w_{\vR1}&=\left( \frac{\sqrt{2+\delta_+^2}}{\sqrt{2+\delta^2}},-\frac{e^{3i\pi/4}}{\sqrt{2+\delta_+^2}},0,\frac{e^{i\pi/4}}{\sqrt{2+\delta_+^2}} \right)^T,\nonumber\\
	\bar w_{\vR2}&=\left( 0,\frac{\sqrt{1+\frac{\delta_-^2}{2}}}{2\sqrt{2+\delta^2}},\frac{e^{-i\pi/4}}{\sqrt{1+\frac{\delta_-^2}{2}}},\frac{i\sqrt{1+\frac{\delta_-^2}{2}}}{2\sqrt{2+\delta^2}}\right)^T,
	\label{eq:BBHW}
\end{align}
where the basis for the spinor is
$\{\ket{\vR,1},\ket{\vR+\va_x,2},\ket{\vR+\va_x+\va_y,3},\ket{\vR+\va_y,4}\}$,
and the notation $\delta_\pm=\delta\pm\sqrt{2+\delta^2}$ is used. The above
bulk Wannier functions give the following components of the bulk quadrupole
tensor~(\ref{eq:hatq})
\begin{align}
	\hat q_{xy}&=\frac{e}{1+\delta_-^2/2}\mod e,\nonumber\\
	\hat q_{xx}&=\hat q_{yy}=0\mod e.
	\label{eq:BBHq}
\end{align}
\begin{figure}[t]
	\centering
	\includegraphics[width=.8\columnwidth]{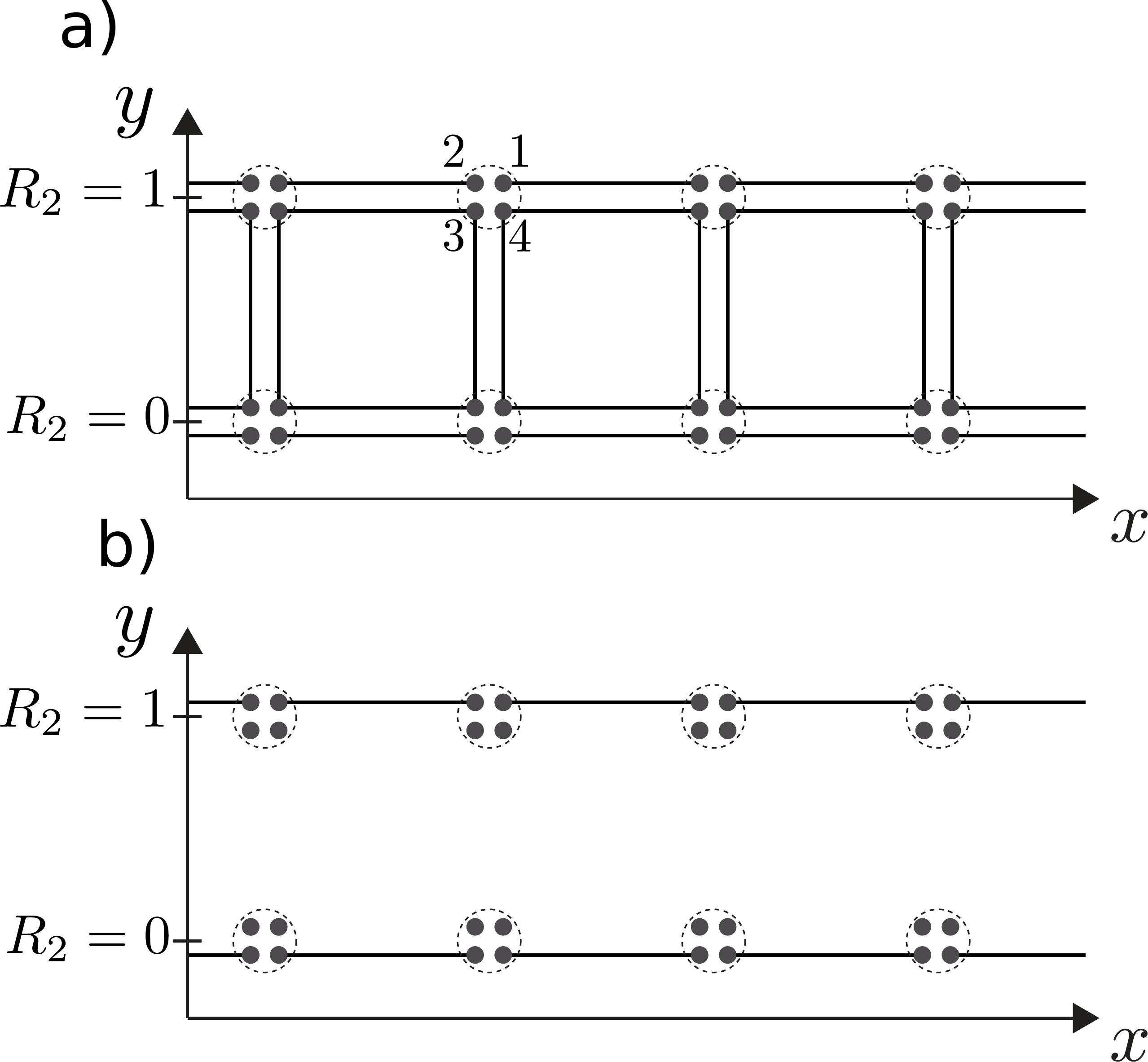}
	\caption{The ribbon corresponding to the BBH model~(\ref{eq:BBH}) with
	theorist boundary conditions in the $y$ direction. The bulk Wannier
	functions can be chosen to be localized on four sites and are
	represented by the squares (a).  The projector ${\cal P}^{L=1}_{k_x}$
	is obtain by removing the bulk Wannier functions belonging to the
	unit cell at $R_2=0$ (b); see Eq.~(\ref{eq:17}). Dashed circles group the
	sites that belong to the same unit cell and integers correspond to
	$\gamma$.}
\label{fig:BBH}
\end{figure}

Wannier edge polarization $\vP_x$ is computed by considering ribbon with
theorist boundary conditions in the $y$ direction. We take the ribbon supercell to
consist of two unit cells (eight sites) with coordinates $R_y=0,1$; see
Fig.~\ref{fig:BBH}.  The projector ${\cal P}^{L=1}_{k_x}$ onto the ribbon
states after performing Wannier cut, which removes the hybrid bulk WFs with
$R_2=0$ [see Eq.~(\ref{eq:17})], reads 
\begin{align}
	{\cal P}_{k_x}^{L=1}&=
	\begin{pmatrix}
		\ket{\psi^\text{edge}_{k_x}}\bra{\psi^\text{edge}_{k_x}} & 0_{2\times4} & 0_{2\times2}\\
		0_{4\times2} & 0_{4\times4} & 0_{4\times2}\\
		0_{2\times2} & 0_{2\times4} & \ket{\psi^\text{edge'}_{k_x}}\bra{\psi^\text{edge'}_{k_x}}
	\end{pmatrix},\nonumber\\
	\ket{\psi^\text{edge}_{k_x}}&=\left( \frac{\tilde\delta_+}{\sqrt{1+\tilde\delta_+^2}},-\frac{e^{-i\pi/4 e^{ik_xa}}}{\sqrt{1+\tilde\delta_+^2}} \right)^T,
	\label{eq:BBHPkx}
\end{align}
where $\tilde\delta_+=\delta+\sqrt{1+\delta^2}$ and
$\ket{\psi^\text{edge'}_{k_x}}={\cal C}_4^2 \ket{\psi^\text{edge}_{-k_x}}$; see
Fig.~\ref{fig:BBH}b.  Taking the corresponding subblock of the matrix ${\cal
P}_{k_x}^{L=1}$ [see Eq.~(\ref{eq:cut0})] gives the projector onto the upper
edge subsystem 
\begin{align}
	{\cal P}_{k_x}^{\text{edge}}&=
	\begin{pmatrix}
		\ket{\psi^\text{edge}_{k_x}}\bra{\psi^\text{edge}_{k_x}} & 0_{2\times2}\\
		0_{2\times2} & 0_{2\times2}
	\end{pmatrix}.
	\label{eq:BBHPkxedge}
\end{align}
Repeating the same calculation for the ribbon along the $y$ direction, after
substituting ${\cal P}_{k_x}^{\text{edge}}$ (${\cal P}_{k_y}^{\text{edge}}$)
into Eq.~(\ref{eq:3}), we obtain the two Wannier edge polarizations,
\begin{align}
	\vP^\text{edge}_\alpha&=\frac{ae}{1+\tilde\delta_+^2} \ve_\alpha.
	\label{eq:BBHvPedge}
\end{align}
The edge polarizations~(\ref{eq:Pedge}) read
\begin{align}
	\vec P^\text{edge}_\alpha&=\left( \frac{1}{1+\tilde\delta_+^2}+\frac{1}{2+\delta_-^2} \right)ae\ve_\alpha.
	\label{eq:BBHPedge}
\end{align}
The above result implies that the edge dipoles vanish $D^\text{edge}_\alpha=0$,
while the corner charge of the upper-right corner $Q^\text{c}$ reads
\begin{align}
	Q^\text{c}&=\frac{2e}{1+\tilde\delta_+^2}+\frac{e}{1+\delta_-^2/2}\mod e.
	\label{eq:BBHQc}
\end{align}
Alternatively, considering the flake with theorist boundary conditions in both
$x$ and $y$ directions, the corner charge $Q^\text{c}$ can be computed via
Eq.~(\ref{eq:Qc_ramp}), which agrees with the result~(\ref{eq:BBHQc}).

\subsection{Orbitals without internal quadrupole moment}
This example considers a two-dimensional tight-binding model with two sites per
unit cell, defined on an arbitrary Bravais lattice with primitive vectors $\vec
a_1$ and $\vec a_2$. The Hamiltonian is written as
\begin{align}
	\label{eq:23}
	h=&\sum_{\vec R}\left[ \sum_{\gamma=1,2}\left((-1)^\gamma\delta\vert\vec R\gamma\rangle\langle\vec R\gamma\vert\right.\right.\\
		&\left.\left.+\sum_{d=1,2} ((-1)^\gamma t\vert\vec R\gamma \rangle\langle\vec R+\vec a_d\bar\gamma\vert+ t_\gamma\vert\vec R\gamma\rangle\langle\vec R+\vec a_d \gamma\vert)\right)\right],\nonumber
\end{align}
where $\vert\vec R\gamma\rangle$ is the $\gamma$ orbital at the position $\vec
R$ and $\{\bar1,\bar2\}=\{2,1\}$. The above Hamiltonian has inversion symmetry
that maps the $\gamma$ orbital into itself; hence the bulk polarization is
quantized. We make a choice of parameters of Hamiltonian~(\ref{eq:23}) such
that the bulk is gapped at half-filling and for theorist boundary conditions
the corner charge is sizable, $\delta=-1$, $t=-0.08$, $t_1=3.5\times t$, and
$t_2=-1.5\times t$. It is easy to see that for these parameters, the bulk
polarization vanishes.
\begin{figure}[h]
	\centering
	\includegraphics[width=.9\columnwidth]{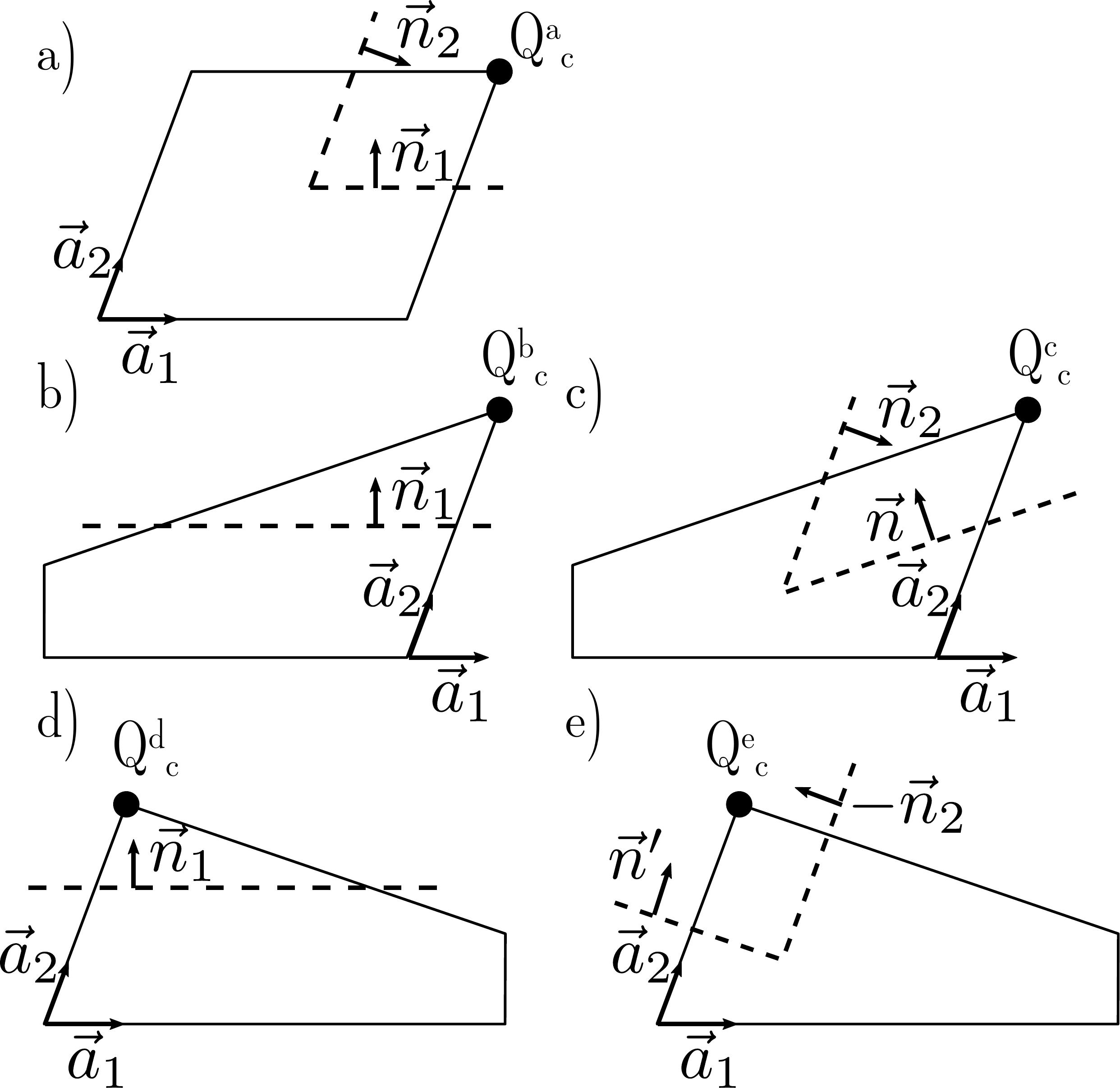}
	\caption{Three different corners for the system described by
		the Hamiltonian~(\ref{eq:23}) with the corresponding corner
		regions (dashed lines).  The corner between the edges along the
		primitive vectors $\vec a_1$ and $\vec a_2$, with the corner
		region parallel to the edges (a). The corner defined by the
		lattice vectors $\vec e_1=2\vec a_1+\vec a_2$ and $\vec a_2$
		(b)-(c). Choosing the corner region above the line along $\vec a_1$
		(b) results in a different corner charge compared to the corner
		region parallel to the edges (c), i.e., $Q^\text{c}_b\neq
		Q^\text{c}_c$. The same as in panels (b) and (c) for the corner between the
		edges along $\vec e_1^\prime=2\vec a_1-\vec a_2$ and $\vec a_2$ is shown in panels
		(d) and (e). The unit normal vectors $\vec n_1$, $\vec n_2$ [$-\vec
		n_2$ for panel (e)], $\vec n$ and $\vec n^\prime$ are oriented
		to point toward the corner.}
	\label{fig:example}
\end{figure}

The occupied bulk WF $\vert\bar w_{\vec R}\rangle$ is chosen as follows. When
all the hoppings are switched off $t=0$, the maximally localized WF takes the
form $\vert\bar w_{\vec R}^0\rangle=\vert\vec R 1\rangle$ for $\delta<0$. The
corresponding (smooth) Bloch eigenfunction $\vert \psi_{\vec k}^0\rangle$ is
given as
\begin{align}
	\vert \psi_{\vec k}^0\rangle&=\langle\tilde\psi_{\vec k}\vert\bar w_{\vec R}^0\rangle\vert\tilde\psi_{\vec k}^0\rangle,
	\label{eq:25}
\end{align}
for $\vert\tilde\psi_{\vec k}^0\rangle$ (not necessarily smooth) Bloch
eigenfunction. A smooth gauge $\vert\psi_{\vec k}^{t}\rangle$ for a finite
value of the parameter $t$ is obtained by parallel transport of
$\vert\psi_{\vec k}^0\rangle$ as the hoppings are switched on
\begin{align}
	\vert\psi_{\vec k}^{t}\rangle&=\langle\tilde\psi^{t}_{\vec k}\vert \left( \prod_{t^\prime=0}^{t}\vert\tilde\psi_{\vec k}^{t^\prime}\rangle\langle\tilde\psi_{\vec k}^{t^\prime}\vert \right)\vert\psi_{\vec k}^0\rangle\vert\tilde\psi_{\vec k}^{t}\rangle.
	\label{eq:26}
\end{align}
We set $t=-0.08$ and drop the superscript $t$ from now on. The bulk WF
takes the form
\begin{align}
	\vert\bar w_{\vec R}\rangle=\sum_{\vec R}e^{i\vec k\cdot\vec R}\vert\psi_{\vec k}\rangle.
	\label{eq:27}
\end{align}
Inspecting the values of $\vert\bar w_{(0,0)}(\vec R)\vert^2$ for different
unit cells, we observe that the obtained WF is well localized, with
$99.5\%$ of the charge lying within the unit cell at $(0,0)$. Using
Eq.~(\ref{eq:hatq}), we obtain the bulk quadrupole tensor $\hat q$ with
components in the $(\vec a_1,\vec a_2)$ basis
\begin{align}
	\label{eq:28a}
	\hat q_{12}&=\hat q_{21}=-1.65919\times 10^{-3}e,\\
	\hat q_{11}&=\hat q_{22}=-7.34447\times 10^{-3}e,
	\label{eq:28b}
\end{align}
where we assumed the ion charge $e$ is localized at $\vec R$.

\textit{Edges along primitive vectors $\vec a_1$ and $\vec a_2$.---}We now
perform a ribbon calculation for a ribbon with periodic boundary conditions
along the $\vec a_1$ direction and the supercell consisting of unit cells at
positions $R_2\in-[N_2,N_2]$, $N_2=20$. The Fourier transform of
Eq.~(\ref{eq:23}) gives Bloch Hamiltonian $h_{k_1}$. Substituting the bulk WF
$\vert\bar w_{\vec R}\rangle$ into Eq.~(\ref{eq:15}), the hybrid bulk WF
$\vert\bar w_{k_1R_2}\rangle$ is obtained. We find that the
projector~(\ref{eq:17}) ${\cal P}_{k_1}^L$ for $L=14$ satisfies the
criteria~(\ref{eq:middle_layer0}) and (\ref{eq:neutrality}) both with accuracy of
$10^{-12}$. From Eq.~(\ref{eq:cut0}), the edge projector onto the top edge
along $\vec a_1$, ${\cal P}^\text{edge}_{k_1}$, is obtained, giving the Wannier
edge polarization
\begin{align}
	\vP^\text{edge}_{\vec a_1}&=(0.17609\vec a_1+3.24851\vec a_2)\times 10^{-3}e,
	\label{eq:30}
\end{align}
where in Eq.~(\ref{eq:3}) we used the grid of $120$ equally spaced
$k_1$ points. [In this section, we index the (Wannier) edge polarizations with
the corresponding edge lattice vector instead of the integer index.] The same
procedure is repeated for the right edge along $\vec a_2$, which gives
$\vP_{\vec a_2}^\text{edge}$ obtained from Eq.~(\ref{eq:30}) after setting
$\vec a_\alpha\rightarrow\vec a_{\bar\alpha}$. Substituting
Eqs.~(\ref{eq:28a})-(\ref{eq:30}) into the expression for the edge
polarization~(\ref{eq:Pedge}) gives
\begin{align}
	\vec P^\text{edge}_{\vec a_\alpha} &=(-0.65350\vec a_\alpha-0.42372\vec a_{\bar\alpha})\times 10^{-3}e.
	\label{eq:ex1Pedge}
\end{align}

\textit{Edge along $\vec e_1=2\vec a_1+\vec a_2$---}We consider a ribbon,
periodic in $\vec e_1=2\vec a_1+\vec a_2$ direction, with supercell along $\vec
a_1$-direction. Repeating the same calculation as above, we obtain for
$N_2=30$, that choosing $L=15$ satisfies conditions~(\ref{eq:middle_layer0}) and
(\ref{eq:neutrality}) both with accuracy of about $10^{-8}$. The Wannier edge
polarization for the upper edge of this ribbon is
\begin{align}
	\vP^\text{edge}_{\vec e_1}&=(6.05756\vec e_1-14.76152\vec a_1)\times 10^{-3}e.
	\label{eq:ex2Wedge}
\end{align}
The above result together with Eqs.~(\ref{eq:Pedge}),
(\ref{eq:28a}), and (\ref{eq:28b}) give the edge polarization $\vec
P^\text{edge}_{\vec e_1}$ for the edge along $\vec e_1$
\begin{align}
	\vec P^\text{edge}_{\vec e_1} &=(-0.45731\vec e_1+0.28122\vec a_1)\times 10^{-3}e.
	\label{eq:ex2Pedge}
\end{align}

\textit{Edge along $\vec e^\prime_1=2\vec a_1-\vec a_2$---}Proceeding as above,
we obtain for the Wannier edge polarization for the upper edge of the ribbon
along $\vec e^\prime_1$
\begin{align}
	\vP^\text{edge}_{\vec e_1^\prime}&=(-6.76134\vec e_1+17.14810\vec a_1)\times 10^{-3}e,
	\label{eq:ex3Wedge}
\end{align}
and the corresponding edge polarization,
\begin{align}
	\vec P^\text{edge}_{\vec e_1^\prime} &=(1.41273\vec e_1-4.52168\vec a_1)\times 10^{-3}e.
	\label{eq:ex4Pedge}
\end{align}

\textit{Edge dipole $D^\text{edge}$---}We confirm that
relation~(\ref{eq:PedgeDedge}) holds exactly. For example, for the edges along
the lattice vectors $\vec a_\alpha$
\begin{align}
	\frac{\vec P^\text{edge}_{\vec a_\alpha}\cdot\vec n_\alpha}{\ve_{\bar\alpha}\cdot\vec n_\alpha}&=-0.42372\times 10^{-3}e,
	\label{eq:exDedge}
\end{align}
agrees with the value obtained from either flake or ribbon calculation
$D^\text{edge}_{\vec a_\alpha}/\ve_{\bar\alpha}\cdot\vec
n_\alpha=-0.42372\times 10^{-3}e$.

\textit{Corner charge.---}Three different corners, as shown in
Fig.~\ref{fig:example}, are considered.  To compute the corner charge, we
consider a flake with the boundaries along the edge vectors, with the
lower-left corner located at $-N_1\vec e_1-N_2\vec e_2$ and the upper-right
corner at $N_1\vec e_1+N_2\vec e_2$.  After diagonalization of the
Hamiltonian~(\ref{eq:23}) for $N_1=N_2=20$, the charge density $\rho(\vr)$ is
obtained from Eq.~(\ref{eq:rhor}).

For the situation in Fig.~\ref{fig:example}a, the corner charge $Q_a^\text{c}$
is obtained by integrating the charge density~(\ref{eq:rhor}) over the corner
area denoted by dashed lines; see Eq.~(\ref{eq:Qc_ramp}). We obtain
$Q_a^\text{c}=-1.30687\times10^{-3}e$ which should be compared with
Eq.~(\ref{eq:bcc}),
\begin{align}
	Q^\text{c}_a&=\frac{L_2}{A_\text{cell}} \vec P^\text{edge}_{\vec a_1}\cdot\vec n_2+\frac{L_1}{A_\text{cell}}\vec P^\text{edge}_{\vec a_2}\cdot \vec n_1\nonumber\\
	&=-1.30700\times10^{-3}e,
	\label{eq:Qca}
\end{align}
where $L_\alpha=\vert \vec a_\alpha\vert$, and $\vec n_\alpha$ is a unit normal
vector as depicted in Fig.~\ref{fig:example}a.

Figures~\ref{fig:example}b and c consider the corner formed by the edges $\vec
e_1=2\vec a_1+\vec a_2$ and $\vec e_2=\vec a_2$. For the corner region as in
Fig.~\ref{fig:example}b, we plug the microscopic charge density~(\ref{eq:rhor})
into Eq.~(\ref{eq:rho_m}) and use Gaussian function. The integration of
macroscopic charge density over the corner region yields the corner charge
$Q^\text{c}_b=-1.11059\times10^{-3}e$ that should be compared with
\begin{align}
	Q^\text{c}_b&=\frac{L_1}{A_\text{cell}}\vec P^\text{edge}_{\vec e_1}\cdot\vec n_1+\frac{L_1}{A_\text{cell}}\vec P^\text{edge}_{\vec a_2}\cdot \vec n_1\nonumber\\
	&=-1.11081\times10^{-3}e.
	\label{eq:Qcb}
\end{align}
On the other hand, for the corner region in Fig.~\ref{fig:example}c, the corner
charge $Q^\text{c}_c$ can be obtained from Eq.~(\ref{eq:Qc_ramp}). The result
$Q^\text{c}_c=-0.75998\times10^{-3}e$ agrees well with
\begin{align}
	Q^\text{c}_c&=\frac{L_2}{A_\text{cell}}\vec P^\text{edge}_{\vec e_1}\cdot \vec n_2+\frac{L}{A_\text{cell}}\vec P^\text{edge}_{\vec a_2}\cdot \vec n\nonumber\\
	&=-0.75834\times10^{-3}e,
	\label{eq:Qcc}
\end{align}
where $L=\vert\vec e_1\vert$, and $\vec n$ is the unit normal vector as shown in
Fig.~\ref{fig:example}c.

The third corner that we consider is shown in Figs.~\ref{fig:example}d and e, formed
by the lattice vectors $\vec e_1^\prime=2\vec a_1-\vec a_2$ and $\vec a_2$. For
the corner region defined by the line along $\vec a_1$ [see
Fig.~\ref{fig:example}d], the corner charge is
$Q_d^\text{c}=-0.75806\times10^{-3}e$. On the other hand, from bulk-and-edge to
corner charge correspondence~(\ref{eq:bcc}), we obtain
\begin{align}
	Q^\text{c}_d&=\frac{L_1}{A_\text{cell}}\vec P^\text{edge}_{\vec e_1^\prime}\cdot \vec n_1+\frac{L_1}{A_\text{cell}}(-\vec P^\text{edge}_{\vec a_2})\cdot \vec n_1\nonumber\\
	&=-0.75923\times10^{-3}e,
	\label{eq:Qcd}
\end{align}
where we used that the edge polarization for the left edge along $\vec a_2$ in
Fig.~\ref{fig:example}a is minus that of the right edge, i.e.,
$-P^\text{edge}_{\vec a_2}$. This relation holds for the present example
because the system is inversion symmetric. Finally, for the corner region in
Fig.~\ref{fig:example}e, we use Eq.~(\ref{eq:Qc_ramp}) to obtain the corner
charge $Q^\text{c}_e=1.71431\times10^{-3}e$ that agrees with
\begin{align}
	Q^\text{c}_e&=\frac{L^\prime}{A_\text{cell}}(-\vec P^\text{edge}_{\vec a_2^\prime})\cdot \vec n^\prime+\frac{L_2}{A_\text{cell}}\vec P^\text{edge}_{\vec e_1^\prime}\cdot (-\vec n_2)\nonumber\\
	&=1.71347\times10^{-3}e.
	\label{eq:Qce}
\end{align}
In the above expression, we used the notation $L^\prime=\vert\vec e_1^\prime\vert$,
and the unit-normal vector $\vec n^\prime$ is shown in Fig.~\ref{fig:example}e.

\textit{Flake's quadrupole moment tensor.---}For comparison, we also compute
the quadrupole tensor for the inversion symmetric flake shown in
Fig.~\ref{fig:example}a
\begin{align}
	\hat q_{12}^\text{flake}&=\hat q_{21}^\text{flake}=-1.1582\times 10^{-3}e,\\
	\hat q_{11}^\text{flake}&=\hat q_{22}^\text{flake}=-7.43307\times 10^{-4}e.
	\label{eq:34}
\end{align}
We observe that $\hat q_{12}^\text{flake}$ does not agree well with the corner
charge $Q^\text{c}_a$ for the flake size $N_1=N_2=20$; see the last paragraph
of Sec.~\ref{ssec:discussion}.

\subsection{Orbitals with internal quadrupole moment}
In the previous example, we assumed that the orbitals of the tight-binding
model~(\ref{eq:23}) are isotropic, and hence they themselves have vanishing
quadrupole moment with respect to their center of mass. To include possible
quadrupole moments of the electron orbitals, we can replace the corresponding
$\delta$ function for the electrons in Eq.~(\ref{eq:rhor}) with the actual
shape of the electron-orbital's charge density. Alternatively, we can perform
following unitary transformation to the Hamiltonian~(\ref{eq:23}) which changes
the basis from $\vert\vec R\gamma\rangle$ to $\vert\vec R\tilde\gamma\rangle$,
\begin{align}
	\vert\vec R 1\rangle&=\frac{1}{\sqrt{2}}(\vert\vec R\tilde 1\rangle+\vert\vec R\tilde 2\rangle),\\
	\vert\vec R 2\rangle&=\frac{1}{\sqrt{2}}(\vert\vec R\tilde 1\rangle-\vert\vec R\tilde 2\rangle).
	\label{eq:35}
\end{align}
Since inversion symmetry maps the two new orbitals as $\tilde
1\leftrightarrow\tilde 2$, one can move the orbitals $\vert\vec R\tilde
\gamma\rangle$ to the positions $\vec R+(-1)^{\tilde\gamma}\vec X$, where $\vec
X=X_1\vec a_1+X_2\vec a_2$ with $X_1,X_2\in[-1/2,1/2]$. The obtained
tight-binding model is the same as the model~(\ref{eq:23}), although it has
quadrupole moment tensor with components in reduced coordinates equal to $\hat
q_{12}=\hat q_{21}=-eX_1X_2$, $\hat q_{11}=-eX_1^2$, and $\hat q_{22}=-eX_2^2$
for the case when all the hoppings are set to zero.

In order to calculate the corner charge $Q^\text{c}$ of a flake for the
top right corner in Fig.~\ref{fig:example}a, we proceed as in the previous
section, where the charge density~(\ref{eq:rhor}) is changed compared to the
previous example because the positions of the orbitals changed:
\begin{align}
	\rho(\vr)=-e\sum_{\vec R n \tilde\gamma}&\left(\vert\langle\vec R\tilde\gamma\vert\psi_n\rangle\vert^2\delta(\vr-\vec R-(-1)^{\tilde\gamma}\vec X)\right.\nonumber\\
	&\left.-\delta(\vr-\vec R)\right).
	\label{eq:36}
\end{align}
We obtain the corner charge of the flake by substituting the above charge
density into Eq.~(\ref{eq:Qc_ramp}),
\begin{align}
	Q^\text{c}&=Q^\text{c}_{a}+0.0101783e\times(X_1+X_2)-eX_1X_2,
	\label{eq:37}
\end{align}
where $Q^\text{c}_{a}$ is the corner charge~(\ref{eq:Qca}) for the case
when both the orbitals are placed at the position $\vec R$, i.e., $\vec X=0$.
Similarly, the calculation of the bulk quadrupole tensor, using the bulk WF
from the previous example, gives
\begin{align}
	\label{eq:38a}
	\hat q_{12}&=\hat q_{21}=\hat q_{12}^{(0,0)}-eX_1X_2,\\
	\label{eq:38b}
	\hat q_{11}&=\hat q_{11}^{(0,0)}-eX_1^2,\\
	\label{eq:38c}
	\hat q_{22}&=\hat q_{22}^{(0,0)}-eX_2^2,
\end{align}
where the superscript ``$(0,0)$'' denotes the quadrupole moment tensor in
Eqs.~(\ref{eq:28a}) and (\ref{eq:28b}). The calculation of Wannier edge
polarization has additional contribution from the second term in
Eq.~(\ref{eq:3}),
\begin{align}
	\vP_{\vec a_\alpha}^{\text{edge}}&=\vP_{\vec a_\alpha}^{\text{edge},(0,0)}+0.0101784e\times \vec X,
	\label{eq:39}
\end{align}
where the term $\vP_{\vec a_\alpha}^{\text{edge},(0,0)}$ is given by
Eq.~(\ref{eq:30}). Similarly, the edge polarization~(\ref{eq:ex1Pedge}) gets
modified to
\begin{align}
	\vec P_{\vec a_\alpha}^{\text{edge}}=&\vec P_{\vec a_\alpha}^{\text{edge},(0,0)}+0.0101784e\times \vec X-\frac{1}{2}e X_{\bar\alpha}^2\vec a_{\bar\alpha}\nonumber\\
	&-\frac{1}{2}eX_1X_2\vec a_{\alpha}.
	\label{eq:exfPedge}
\end{align}
The bulk-and-edge to corner charge correspondence~(\ref{eq:bcc}) gives the
corner charge
\begin{align}
	\label{eq:exf}
	Q^\text{c}&=\frac{L_2}{A_\text{cell}} \vec P^\text{edge}_{\vec a_1}\cdot\vec n_2+\frac{L_1}{A_\text{cell}}\vec P^\text{edge}_{\vec a_2}\cdot \vec n_1\\
	&=Q^\text{c}_a+0.0101784e\times(X_1+X_2)-eX_1X_2\nonumber,
\end{align}
which agrees well with the independent corner charge calculation~(\ref{eq:37}).

Note that the model studied in this example for $X_1=X_2=1/6$ is the same
as the model previously studied in Ref.~\onlinecite{zhou2015}, although here we
chose different hopping parameters. We are now in position to understand why
the contribution from the bulk quadrupole tensor $\hat q$ was previously
overlooked.~\cite{zhou2015} Reference~\onlinecite{zhou2015} assumes that $e/2$
ionic charge is localized at each lattice site instead of ionic charge
$e$ localized at $\vec R$ as was done in Eq.~(\ref{eq:36}). Therefore, the
charge density~(\ref{eq:36}) is modified to 
\begin{align}
	\rho(\vr)=-e\sum_{\vec R n \tilde\gamma}&\left(\vert\langle\vec R\tilde\gamma\vert\psi_n\rangle\vert^2-\frac{1}{2}\right)\delta(\vr-\vec R-(-1)^{\tilde\gamma}\vec X).
	\label{eq:41}
\end{align}
The resulting corner charge $Q^\text{c}$ is given by Eq.~(\ref{eq:37}) with the
last term $-eX_1X_2$ omitted. The above modification of the ionic charge
density changes the expressions for the quadrupole tensor $\hat q$ in
Eqs.~(\ref{eq:38a})-(\ref{eq:38c}) to $\hat q^{(0,0)}$. Therefore,
Eq.~(\ref{eq:exfPedge}) becomes
\begin{align}
	\vec P_{\vec a_\alpha}^{\text{edge}}=&\vec P_{\vec a_\alpha}^{\text{edge},(0,0)}+0.0101784e\times \vec X,
	\label{eq:Xedge}
\end{align}
where the dominant contribution is from the $\vec X$-dependent piece of the
corresponding Wannier edge polarization~(\ref{eq:39}). For the hopping
parameters chosen in this example, the bulk contribution $\hat q_{12}^{(0,0)}$
from $\vec P_{\vec a_\alpha}^{\text{edge},(0,0)}$ cannot be overlooked for
$X_1=X_2=1/6$. On the other hand, for the hopping parameters used in
Ref.~\onlinecite{zhou2015}, the bulk contribution $\hat q_{12}^{(0,0)}$ is an
order of magnitude smaller than the value in Eq.~(\ref{eq:28a})---easy to
overlook in the presence of $\vec X$-dependent term in
Eq.~(\ref{eq:Xedge}). This statement should be compared to
Eq.~(\ref{eq:exf}), where $\vec X$-dependent bulk contribution $-eX_1X_2$
dominates.

\section{Conclusions}\label{sec:conclusions}
The statement of the bulk-boundary correspondence, formulated by the modern theory of
electrical polarization,~\cite{king-smith1993,vanderbilt1993,vanderbilt2018} is
that the non-vanishing bulk polarization of a two-dimensional insulator
determines the edge charge density. On the other hand, for vanishing bulk
polarization one can still observe boundary signatures in form of corner
charges and edge dipoles. In this work, we prove that corner charges and edge
dipoles of band insulators can be obtained by representing a terminated
crystal as a collection of edge regions with polarization $\vec
P^\text{edge}_\alpha$, where $\alpha$ enumerates the edges (see
Fig.~\ref{fig:1}). We find that the edge polarization $\vec
P^\text{edge}_\alpha$ consists of two pieces, the bulk piece given by the
quadrupole tensor of bulk Wannier functions' charge density, and the edge piece
that we call Wannier edge polarization $\vP^\text{edge}_\alpha$. The Wannier
edge polarization is defined as
polarization of the edge-subsystem, which is obtained by cutting out the region
around the corresponding edge using ``Wannier cut'', the cut that utilizes the
bulk Wannier functions as ``shape cutter''. Within our representation of the
terminated crystal, the edge polarizations $\vec P^\text{edge}_\alpha$
determine the corner charges via mentioned bulk-boundary correspondence. Since
$\vec P^\text{edge}_\alpha$ has both bulk and edge piece, the resulting
correspondence~(\ref{eq:bcc}) is dubbed bulk-and-edge to corner correspondence,
which is the main result of our work. The edge polarizations $\vec
P^\text{edge}_\alpha$ defined in this work depend on the choice of occupied
bulk Wannier functions, which is consistent with the fact that the number of
physical observables (i.e. corner charges and edge dipoles) is one less than
the number of independent components of all edge polarizations.

In the context of this work, only the corner charges and the edge dipole moments are
considered as relevant physical observables characterizing the macroscopic
charge density of a terminated crystal. For this reason, only the lowest
non-vanishing multipole moment, polarization in this case, is taken into
account for the edge regions in Fig.~\ref{fig:1}. For example, one can represent
the crystal as a collection of edge regions that have not only polarization but
also quadrupole moment. Although the quadrupole moment of the edge region
affects neither the corner charge nor the edge dipole (see Fig.~\ref{fig:3}), it
does affect finer (higher order) features~\footnote{In the same spirit, the
corner charges and the edge dipoles are finer (higher-order) observables
compared to the edge charge density.} of the flake's macroscopic charge
density.  Admittedly, even the measurement of corner chargers and edge dipoles may
prove to be experimentally challenging. Here we imagine a setup consisting of a
crystal with $n$ edges, where the application of strain causes the edge
polarizations~(\ref{eq:Pedge}) to change. The change in the edge polarizations
results in the current-flow along the corresponding edge, which can be in
principle measured. The measurement of the currents related to the change of
the edge dipoles requires more local current probes which poses an additional
difficulty.

In this work, we considered two-dimensional systems and we expect the extension
to three-dimensional systems to follow along similar lines. Namely, for
three-dimensional crystal with vanishing bulk polarization, one can consider
hinge charge densities, or if the hinge charge densities vanish, the
corner charges. In the former case, the aim would be to represent the terminated
crystal as a collection of polarized surface regions, whereas in the latter case
one would have polarized hinge regions. Another interesting question in this
context is finding all the symmetry constraints that quantize some of the
mentioned boundary signatures, where the bulk Wannier functions can be chosen
to respect the symmetry constraint. On a more challenging side, we question
whether the notion of Wannier cut and Wannier edge polarization can be extended
to the systems lacking band structure or even single-particle description,
since in that case, no obvious generalization of Wannier functions
exists.~\cite{souza2000}

\section{Acknowledgements}
I thank Ren Shang, David Vanderbilt, and Haruki Watanabe for useful
discussions. During work on this manuscript, I became aware of a related
(unpublished) work by R.~Shang, I.~Souza, and D.~Vanderbilt that was presented
at the online workshop ``Recent Developments on Multipole Moments in Quantum
Systems''.~\footnote{\url{https://sites.google.com/g.ecc.u-tokyo.ac.jp/workshop-multipole/}}
I acknowledge financial support from the FNS/SNF Ambizione Grant
No.~PZ00P2\_179962.

\bibliography{ref}

\begin{thebibliography}{38}%
\makeatletter
\providecommand \@ifxundefined [1]{%
 \@ifx{#1\undefined}
}%
\providecommand \@ifnum [1]{%
 \ifnum #1\expandafter \@firstoftwo
 \else \expandafter \@secondoftwo
 \fi
}%
\providecommand \@ifx [1]{%
 \ifx #1\expandafter \@firstoftwo
 \else \expandafter \@secondoftwo
 \fi
}%
\providecommand \natexlab [1]{#1}%
\providecommand \enquote  [1]{``#1''}%
\providecommand \bibnamefont  [1]{#1}%
\providecommand \bibfnamefont [1]{#1}%
\providecommand \citenamefont [1]{#1}%
\providecommand \href@noop [0]{\@secondoftwo}%
\providecommand \href [0]{\begingroup \@sanitize@url \@href}%
\providecommand \@href[1]{\@@startlink{#1}\@@href}%
\providecommand \@@href[1]{\endgroup#1\@@endlink}%
\providecommand \@sanitize@url [0]{\catcode `\\12\catcode `\$12\catcode
  `\&12\catcode `\#12\catcode `\^12\catcode `\_12\catcode `\%12\relax}%
\providecommand \@@startlink[1]{}%
\providecommand \@@endlink[0]{}%
\providecommand \url  [0]{\begingroup\@sanitize@url \@url }%
\providecommand \@url [1]{\endgroup\@href {#1}{\urlprefix }}%
\providecommand \urlprefix  [0]{URL }%
\providecommand \Eprint [0]{\href }%
\providecommand \doibase [0]{http://dx.doi.org/}%
\providecommand \selectlanguage [0]{\@gobble}%
\providecommand \bibinfo  [0]{\@secondoftwo}%
\providecommand \bibfield  [0]{\@secondoftwo}%
\providecommand \translation [1]{[#1]}%
\providecommand \BibitemOpen [0]{}%
\providecommand \bibitemStop [0]{}%
\providecommand \bibitemNoStop [0]{.\EOS\space}%
\providecommand \EOS [0]{\spacefactor3000\relax}%
\providecommand \BibitemShut  [1]{\csname bibitem#1\endcsname}%
\let\auto@bib@innerbib\@empty
\bibitem [{\citenamefont {Resta}\ and\ \citenamefont
  {Vanderbilt}(2007)}]{resta2007}%
  \BibitemOpen
  \bibfield  {author} {\bibinfo {author} {\bibfnamefont {R.}~\bibnamefont
  {Resta}}\ and\ \bibinfo {author} {\bibfnamefont {D.}~\bibnamefont
  {Vanderbilt}},\ }\href {\doibase 10.1007/978-3-540-34591-6_2} {\emph
  {\bibinfo {title} {Physics of Ferroelectrics: A Modern Perspective}}}\
  (\bibinfo  {publisher} {Springer Berlin Heidelberg},\ \bibinfo {address}
  {Berlin, Heidelberg},\ \bibinfo {year} {2007})\ pp.\ \bibinfo {pages}
  {31--68}\BibitemShut {NoStop}%
\bibitem [{\citenamefont {King-Smith}\ and\ \citenamefont
  {Vanderbilt}(1993)}]{king-smith1993}%
  \BibitemOpen
  \bibfield  {author} {\bibinfo {author} {\bibfnamefont {R.~D.}\ \bibnamefont
  {King-Smith}}\ and\ \bibinfo {author} {\bibfnamefont {D.}~\bibnamefont
  {Vanderbilt}},\ }\href {\doibase 10.1103/PhysRevB.47.1651} {\bibfield
  {journal} {\bibinfo  {journal} {Phys. Rev. B}\ }\textbf {\bibinfo {volume}
  {47}},\ \bibinfo {pages} {1651} (\bibinfo {year} {1993})}\BibitemShut
  {NoStop}%
\bibitem [{\citenamefont {Vanderbilt}\ and\ \citenamefont
  {King-Smith}(1993)}]{vanderbilt1993}%
  \BibitemOpen
  \bibfield  {author} {\bibinfo {author} {\bibfnamefont {D.}~\bibnamefont
  {Vanderbilt}}\ and\ \bibinfo {author} {\bibfnamefont {R.~D.}\ \bibnamefont
  {King-Smith}},\ }\href {\doibase 10.1103/PhysRevB.48.4442} {\bibfield
  {journal} {\bibinfo  {journal} {Phys. Rev. B}\ }\textbf {\bibinfo {volume}
  {48}},\ \bibinfo {pages} {4442} (\bibinfo {year} {1993})}\BibitemShut
  {NoStop}%
\bibitem [{\citenamefont {Resta}(1994)}]{resta1994}%
  \BibitemOpen
  \bibfield  {author} {\bibinfo {author} {\bibfnamefont {R.}~\bibnamefont
  {Resta}},\ }\href {\doibase 10.1103/RevModPhys.66.899} {\bibfield  {journal}
  {\bibinfo  {journal} {Rev. Mod. Phys.}\ }\textbf {\bibinfo {volume} {66}},\
  \bibinfo {pages} {899} (\bibinfo {year} {1994})}\BibitemShut {NoStop}%
\bibitem [{\citenamefont {Resta}\ and\ \citenamefont
  {Sorella}(1999)}]{resta1999}%
  \BibitemOpen
  \bibfield  {author} {\bibinfo {author} {\bibfnamefont {R.}~\bibnamefont
  {Resta}}\ and\ \bibinfo {author} {\bibfnamefont {S.}~\bibnamefont
  {Sorella}},\ }\href {\doibase 10.1103/PhysRevLett.82.370} {\bibfield
  {journal} {\bibinfo  {journal} {Phys. Rev. Lett.}\ }\textbf {\bibinfo
  {volume} {82}},\ \bibinfo {pages} {370} (\bibinfo {year} {1999})}\BibitemShut
  {NoStop}%
\bibitem [{\citenamefont {Thonhauser}\ \emph {et~al.}(2005)\citenamefont
  {Thonhauser}, \citenamefont {Ceresoli}, \citenamefont {Vanderbilt},\ and\
  \citenamefont {Resta}}]{thonhauser2005}%
  \BibitemOpen
  \bibfield  {author} {\bibinfo {author} {\bibfnamefont {T.}~\bibnamefont
  {Thonhauser}}, \bibinfo {author} {\bibfnamefont {D.}~\bibnamefont
  {Ceresoli}}, \bibinfo {author} {\bibfnamefont {D.}~\bibnamefont
  {Vanderbilt}}, \ and\ \bibinfo {author} {\bibfnamefont {R.}~\bibnamefont
  {Resta}},\ }\href {\doibase 10.1103/PhysRevLett.95.137205} {\bibfield
  {journal} {\bibinfo  {journal} {Phys. Rev. Lett.}\ }\textbf {\bibinfo
  {volume} {95}},\ \bibinfo {pages} {137205} (\bibinfo {year}
  {2005})}\BibitemShut {NoStop}%
\bibitem [{\citenamefont {Shi}\ \emph {et~al.}(2007)\citenamefont {Shi},
  \citenamefont {Vignale}, \citenamefont {Xiao},\ and\ \citenamefont
  {Niu}}]{shi2007}%
  \BibitemOpen
  \bibfield  {author} {\bibinfo {author} {\bibfnamefont {J.}~\bibnamefont
  {Shi}}, \bibinfo {author} {\bibfnamefont {G.}~\bibnamefont {Vignale}},
  \bibinfo {author} {\bibfnamefont {D.}~\bibnamefont {Xiao}}, \ and\ \bibinfo
  {author} {\bibfnamefont {Q.}~\bibnamefont {Niu}},\ }\href {\doibase
  10.1103/PhysRevLett.99.197202} {\bibfield  {journal} {\bibinfo  {journal}
  {Phys. Rev. Lett.}\ }\textbf {\bibinfo {volume} {99}},\ \bibinfo {pages}
  {197202} (\bibinfo {year} {2007})}\BibitemShut {NoStop}%
\bibitem [{\citenamefont {Trifunovic}\ \emph {et~al.}(2019)\citenamefont
  {Trifunovic}, \citenamefont {Ono},\ and\ \citenamefont
  {Watanabe}}]{trifunovic2019b}%
  \BibitemOpen
  \bibfield  {author} {\bibinfo {author} {\bibfnamefont {L.}~\bibnamefont
  {Trifunovic}}, \bibinfo {author} {\bibfnamefont {S.}~\bibnamefont {Ono}}, \
  and\ \bibinfo {author} {\bibfnamefont {H.}~\bibnamefont {Watanabe}},\ }\href
  {\doibase 10.1103/PhysRevB.100.054408} {\bibfield  {journal} {\bibinfo
  {journal} {Phys. Rev. B}\ }\textbf {\bibinfo {volume} {100}},\ \bibinfo
  {pages} {054408} (\bibinfo {year} {2019})}\BibitemShut {NoStop}%
\bibitem [{\citenamefont {Qi}\ \emph {et~al.}(2008)\citenamefont {Qi},
  \citenamefont {Hughes},\ and\ \citenamefont {Zhang}}]{qi2008}%
  \BibitemOpen
  \bibfield  {author} {\bibinfo {author} {\bibfnamefont {X.-L.}\ \bibnamefont
  {Qi}}, \bibinfo {author} {\bibfnamefont {T.~L.}\ \bibnamefont {Hughes}}, \
  and\ \bibinfo {author} {\bibfnamefont {S.-C.}\ \bibnamefont {Zhang}},\ }\href
  {\doibase 10.1103/PhysRevB.78.195424} {\bibfield  {journal} {\bibinfo
  {journal} {Phys. Rev. B}\ }\textbf {\bibinfo {volume} {78}},\ \bibinfo
  {pages} {195424} (\bibinfo {year} {2008})}\BibitemShut {NoStop}%
\bibitem [{\citenamefont {Essin}\ \emph {et~al.}(2009)\citenamefont {Essin},
  \citenamefont {Moore},\ and\ \citenamefont {Vanderbilt}}]{essin2009}%
  \BibitemOpen
  \bibfield  {author} {\bibinfo {author} {\bibfnamefont {A.~M.}\ \bibnamefont
  {Essin}}, \bibinfo {author} {\bibfnamefont {J.~E.}\ \bibnamefont {Moore}}, \
  and\ \bibinfo {author} {\bibfnamefont {D.}~\bibnamefont {Vanderbilt}},\
  }\href {\doibase 10.1103/PhysRevLett.102.146805} {\bibfield  {journal}
  {\bibinfo  {journal} {Phys. Rev. Lett.}\ }\textbf {\bibinfo {volume} {102}},\
  \bibinfo {pages} {146805} (\bibinfo {year} {2009})}\BibitemShut {NoStop}%
\bibitem [{\citenamefont {Kitaev}(2009)}]{kitaev2009}%
  \BibitemOpen
  \bibfield  {author} {\bibinfo {author} {\bibfnamefont {A.}~\bibnamefont
  {Kitaev}},\ }\href {\doibase 10.1063/1.3149495} {\bibfield  {journal}
  {\bibinfo  {journal} {AIP Conference Proceedings}\ }\textbf {\bibinfo
  {volume} {1134}},\ \bibinfo {pages} {22} (\bibinfo {year}
  {2009})}\BibitemShut {NoStop}%
\bibitem [{\citenamefont {Schnyder}\ \emph {et~al.}(2009)\citenamefont
  {Schnyder}, \citenamefont {Ryu}, \citenamefont {Furusaki},\ and\
  \citenamefont {Ludwig}}]{schnyder2009}%
  \BibitemOpen
  \bibfield  {author} {\bibinfo {author} {\bibfnamefont {A.~P.}\ \bibnamefont
  {Schnyder}}, \bibinfo {author} {\bibfnamefont {S.}~\bibnamefont {Ryu}},
  \bibinfo {author} {\bibfnamefont {A.}~\bibnamefont {Furusaki}}, \ and\
  \bibinfo {author} {\bibfnamefont {A.~W.~W.}\ \bibnamefont {Ludwig}},\ }\href
  {\doibase 10.1063/1.3149481} {\bibfield  {journal} {\bibinfo  {journal} {AIP
  Conference Proceedings}\ }\textbf {\bibinfo {volume} {1134}},\ \bibinfo
  {pages} {10} (\bibinfo {year} {2009})}\BibitemShut {NoStop}%
\bibitem [{\citenamefont {Kane}\ and\ \citenamefont
  {Mele}(2005{\natexlab{a}})}]{kane2005b}%
  \BibitemOpen
  \bibfield  {author} {\bibinfo {author} {\bibfnamefont {C.~L.}\ \bibnamefont
  {Kane}}\ and\ \bibinfo {author} {\bibfnamefont {E.~J.}\ \bibnamefont
  {Mele}},\ }\href {\doibase 10.1103/PhysRevLett.95.146802} {\bibfield
  {journal} {\bibinfo  {journal} {Phys. Rev. Lett.}\ }\textbf {\bibinfo
  {volume} {95}},\ \bibinfo {pages} {146802} (\bibinfo {year}
  {2005}{\natexlab{a}})}\BibitemShut {NoStop}%
\bibitem [{\citenamefont {Thouless}\ \emph {et~al.}(1982)\citenamefont
  {Thouless}, \citenamefont {Kohmoto}, \citenamefont {Nightingale},\ and\
  \citenamefont {den Nijs}}]{thouless1982}%
  \BibitemOpen
  \bibfield  {author} {\bibinfo {author} {\bibfnamefont {D.~J.}\ \bibnamefont
  {Thouless}}, \bibinfo {author} {\bibfnamefont {M.}~\bibnamefont {Kohmoto}},
  \bibinfo {author} {\bibfnamefont {M.~P.}\ \bibnamefont {Nightingale}}, \ and\
  \bibinfo {author} {\bibfnamefont {M.}~\bibnamefont {den Nijs}},\ }\href
  {\doibase 10.1103/PhysRevLett.49.405} {\bibfield  {journal} {\bibinfo
  {journal} {Phys. Rev. Lett.}\ }\textbf {\bibinfo {volume} {49}},\ \bibinfo
  {pages} {405} (\bibinfo {year} {1982})}\BibitemShut {NoStop}%
\bibitem [{\citenamefont {Kane}\ and\ \citenamefont
  {Mele}(2005{\natexlab{b}})}]{kane2005}%
  \BibitemOpen
  \bibfield  {author} {\bibinfo {author} {\bibfnamefont {C.~L.}\ \bibnamefont
  {Kane}}\ and\ \bibinfo {author} {\bibfnamefont {E.~J.}\ \bibnamefont
  {Mele}},\ }\href {\doibase 10.1103/PhysRevLett.95.226801} {\bibfield
  {journal} {\bibinfo  {journal} {Phys. Rev. Lett.}\ }\textbf {\bibinfo
  {volume} {95}},\ \bibinfo {pages} {226801} (\bibinfo {year}
  {2005}{\natexlab{b}})}\BibitemShut {NoStop}%
\bibitem [{\citenamefont {Bernevig}\ \emph {et~al.}(2006)\citenamefont
  {Bernevig}, \citenamefont {Hughes},\ and\ \citenamefont
  {Zhang}}]{bernevig2006}%
  \BibitemOpen
  \bibfield  {author} {\bibinfo {author} {\bibfnamefont {B.~A.}\ \bibnamefont
  {Bernevig}}, \bibinfo {author} {\bibfnamefont {T.~L.}\ \bibnamefont
  {Hughes}}, \ and\ \bibinfo {author} {\bibfnamefont {S.-C.}\ \bibnamefont
  {Zhang}},\ }\href {\doibase 10.1126/science.1133734} {\bibfield  {journal}
  {\bibinfo  {journal} {Science}\ }\textbf {\bibinfo {volume} {314}},\ \bibinfo
  {pages} {1757} (\bibinfo {year} {2006})}\BibitemShut {NoStop}%
\bibitem [{\citenamefont {Kitaev}(2001)}]{kitaev2001}%
  \BibitemOpen
  \bibfield  {author} {\bibinfo {author} {\bibfnamefont {A.~Y.}\ \bibnamefont
  {Kitaev}},\ }\href@noop {} {\bibfield  {journal} {\bibinfo  {journal} {Phys.
  Usp.}\ }\textbf {\bibinfo {volume} {44}},\ \bibinfo {pages} {131} (\bibinfo
  {year} {2001})}\BibitemShut {NoStop}%
\bibitem [{\citenamefont {Alicea}\ \emph {et~al.}(2011)\citenamefont {Alicea},
  \citenamefont {Oreg}, \citenamefont {Refael}, \citenamefont {\mbox{von
  Oppen}},\ and\ \citenamefont {Fisher}}]{alicea2011}%
  \BibitemOpen
  \bibfield  {author} {\bibinfo {author} {\bibfnamefont {J.}~\bibnamefont
  {Alicea}}, \bibinfo {author} {\bibfnamefont {Y.}~\bibnamefont {Oreg}},
  \bibinfo {author} {\bibfnamefont {G.}~\bibnamefont {Refael}}, \bibinfo
  {author} {\bibfnamefont {F.}~\bibnamefont {\mbox{von Oppen}}}, \ and\
  \bibinfo {author} {\bibfnamefont {M.~P.~A.}\ \bibnamefont {Fisher}},\
  }\href@noop {} {\bibfield  {journal} {\bibinfo  {journal} {Nature Phys.}\
  }\textbf {\bibinfo {volume} {7}},\ \bibinfo {pages} {412} (\bibinfo {year}
  {2011})}\BibitemShut {NoStop}%
\bibitem [{\citenamefont {Lutchyn}\ \emph {et~al.}(2010)\citenamefont
  {Lutchyn}, \citenamefont {Sau},\ and\ \citenamefont
  {Das~Sarma}}]{lutchyn2010}%
  \BibitemOpen
  \bibfield  {author} {\bibinfo {author} {\bibfnamefont {R.~M.}\ \bibnamefont
  {Lutchyn}}, \bibinfo {author} {\bibfnamefont {J.~D.}\ \bibnamefont {Sau}}, \
  and\ \bibinfo {author} {\bibfnamefont {S.}~\bibnamefont {Das~Sarma}},\ }\href
  {\doibase 10.1103/PhysRevLett.105.077001} {\bibfield  {journal} {\bibinfo
  {journal} {Phys. Rev. Lett.}\ }\textbf {\bibinfo {volume} {105}},\ \bibinfo
  {pages} {077001} (\bibinfo {year} {2010})}\BibitemShut {NoStop}%
\bibitem [{\citenamefont {Mourik}\ \emph {et~al.}(2012)\citenamefont {Mourik},
  \citenamefont {Zuo}, \citenamefont {Frolov}, \citenamefont {Plissard},
  \citenamefont {Bakkers},\ and\ \citenamefont {Kouwenhoven}}]{mourik2012}%
  \BibitemOpen
  \bibfield  {author} {\bibinfo {author} {\bibfnamefont {V.}~\bibnamefont
  {Mourik}}, \bibinfo {author} {\bibfnamefont {K.}~\bibnamefont {Zuo}},
  \bibinfo {author} {\bibfnamefont {S.~M.}\ \bibnamefont {Frolov}}, \bibinfo
  {author} {\bibfnamefont {S.~R.}\ \bibnamefont {Plissard}}, \bibinfo {author}
  {\bibfnamefont {E.~P. A.~M.}\ \bibnamefont {Bakkers}}, \ and\ \bibinfo
  {author} {\bibfnamefont {L.~P.}\ \bibnamefont {Kouwenhoven}},\ }\href
  {\doibase 10.1126/science.1222360} {\bibfield  {journal} {\bibinfo  {journal}
  {Science}\ }\textbf {\bibinfo {volume} {336}},\ \bibinfo {pages} {1003}
  (\bibinfo {year} {2012})}\BibitemShut {NoStop}%
\bibitem [{\citenamefont {{Shiozaki}}\ \emph {et~al.}(2018)\citenamefont
  {{Shiozaki}}, \citenamefont {{Sato}},\ and\ \citenamefont
  {{Gomi}}}]{shiozaki2018}%
  \BibitemOpen
  \bibfield  {author} {\bibinfo {author} {\bibfnamefont {K.}~\bibnamefont
  {{Shiozaki}}}, \bibinfo {author} {\bibfnamefont {M.}~\bibnamefont {{Sato}}},
  \ and\ \bibinfo {author} {\bibfnamefont {K.}~\bibnamefont {{Gomi}}},\
  }\href@noop {} {\bibfield  {journal} {\bibinfo  {journal} {ArXiv e-prints}\ }
  (\bibinfo {year} {2018})},\ \Eprint {http://arxiv.org/abs/1802.06694}
  {arXiv:1802.06694 [cond-mat.str-el]} \BibitemShut {NoStop}%
\bibitem [{\citenamefont {Trifunovic}\ and\ \citenamefont
  {Brouwer}(2019)}]{trifunovic2019}%
  \BibitemOpen
  \bibfield  {author} {\bibinfo {author} {\bibfnamefont {L.}~\bibnamefont
  {Trifunovic}}\ and\ \bibinfo {author} {\bibfnamefont {P.~W.}\ \bibnamefont
  {Brouwer}},\ }\href {\doibase 10.1103/PhysRevX.9.011012} {\bibfield
  {journal} {\bibinfo  {journal} {Phys. Rev. X}\ }\textbf {\bibinfo {volume}
  {9}},\ \bibinfo {pages} {011012} (\bibinfo {year} {2019})}\BibitemShut
  {NoStop}%
\bibitem [{\citenamefont {Roberts}\ \emph {et~al.}(2020)\citenamefont
  {Roberts}, \citenamefont {Behrends},\ and\ \citenamefont
  {B\'eri}}]{roberts2020}%
  \BibitemOpen
  \bibfield  {author} {\bibinfo {author} {\bibfnamefont {E.}~\bibnamefont
  {Roberts}}, \bibinfo {author} {\bibfnamefont {J.}~\bibnamefont {Behrends}}, \
  and\ \bibinfo {author} {\bibfnamefont {B.}~\bibnamefont {B\'eri}},\ }\href
  {\doibase 10.1103/PhysRevB.101.155133} {\bibfield  {journal} {\bibinfo
  {journal} {Phys. Rev. B}\ }\textbf {\bibinfo {volume} {101}},\ \bibinfo
  {pages} {155133} (\bibinfo {year} {2020})}\BibitemShut {NoStop}%
\bibitem [{\citenamefont {Trifunovic}\ and\ \citenamefont
  {Brouwer}(2020)}]{trifunovic2020}%
  \BibitemOpen
  \bibfield  {author} {\bibinfo {author} {\bibfnamefont {L.}~\bibnamefont
  {Trifunovic}}\ and\ \bibinfo {author} {\bibfnamefont {P.~W.}\ \bibnamefont
  {Brouwer}},\ }\href {\doibase 10.1002/pssb.202000090} {\bibfield  {journal}
  {\bibinfo  {journal} {physica status solidi (b)}\ ,\ \bibinfo {pages}
  {2000090}} (\bibinfo {year} {2020})}\BibitemShut {NoStop}%
\bibitem [{\citenamefont {Benalcazar}\ \emph {et~al.}(2017)\citenamefont
  {Benalcazar}, \citenamefont {Bernevig},\ and\ \citenamefont
  {Hughes}}]{benalcazar2017}%
  \BibitemOpen
  \bibfield  {author} {\bibinfo {author} {\bibfnamefont {W.~A.}\ \bibnamefont
  {Benalcazar}}, \bibinfo {author} {\bibfnamefont {B.~A.}\ \bibnamefont
  {Bernevig}}, \ and\ \bibinfo {author} {\bibfnamefont {T.~L.}\ \bibnamefont
  {Hughes}},\ }\href {\doibase 10.1103/PhysRevB.96.245115} {\bibfield
  {journal} {\bibinfo  {journal} {Phys. Rev. B}\ }\textbf {\bibinfo {volume}
  {96}},\ \bibinfo {pages} {245115} (\bibinfo {year} {2017})}\BibitemShut
  {NoStop}%
\bibitem [{\citenamefont {Ono}\ \emph {et~al.}(2019)\citenamefont {Ono},
  \citenamefont {Trifunovic},\ and\ \citenamefont {Watanabe}}]{ono2019}%
  \BibitemOpen
  \bibfield  {author} {\bibinfo {author} {\bibfnamefont {S.}~\bibnamefont
  {Ono}}, \bibinfo {author} {\bibfnamefont {L.}~\bibnamefont {Trifunovic}}, \
  and\ \bibinfo {author} {\bibfnamefont {H.}~\bibnamefont {Watanabe}},\ }\href
  {\doibase 10.1103/PhysRevB.100.245133} {\bibfield  {journal} {\bibinfo
  {journal} {Phys. Rev. B}\ }\textbf {\bibinfo {volume} {100}},\ \bibinfo
  {pages} {245133} (\bibinfo {year} {2019})}\BibitemShut {NoStop}%
\bibitem [{\citenamefont {Zhou}\ \emph {et~al.}(2015)\citenamefont {Zhou},
  \citenamefont {Rabe},\ and\ \citenamefont {Vanderbilt}}]{zhou2015}%
  \BibitemOpen
  \bibfield  {author} {\bibinfo {author} {\bibfnamefont {Y.}~\bibnamefont
  {Zhou}}, \bibinfo {author} {\bibfnamefont {K.~M.}\ \bibnamefont {Rabe}}, \
  and\ \bibinfo {author} {\bibfnamefont {D.}~\bibnamefont {Vanderbilt}},\
  }\href {\doibase 10.1103/PhysRevB.92.041102} {\bibfield  {journal} {\bibinfo
  {journal} {Phys. Rev. B}\ }\textbf {\bibinfo {volume} {92}},\ \bibinfo
  {pages} {041102} (\bibinfo {year} {2015})}\BibitemShut {NoStop}%
\bibitem [{\citenamefont {Marzari}\ \emph {et~al.}(2012)\citenamefont
  {Marzari}, \citenamefont {Mostofi}, \citenamefont {Yates}, \citenamefont
  {Souza},\ and\ \citenamefont {Vanderbilt}}]{marzari2012}%
  \BibitemOpen
  \bibfield  {author} {\bibinfo {author} {\bibfnamefont {N.}~\bibnamefont
  {Marzari}}, \bibinfo {author} {\bibfnamefont {A.~A.}\ \bibnamefont
  {Mostofi}}, \bibinfo {author} {\bibfnamefont {J.~R.}\ \bibnamefont {Yates}},
  \bibinfo {author} {\bibfnamefont {I.}~\bibnamefont {Souza}}, \ and\ \bibinfo
  {author} {\bibfnamefont {D.}~\bibnamefont {Vanderbilt}},\ }\href {\doibase
  10.1103/RevModPhys.84.1419} {\bibfield  {journal} {\bibinfo  {journal} {Rev.
  Mod. Phys.}\ }\textbf {\bibinfo {volume} {84}},\ \bibinfo {pages} {1419}
  (\bibinfo {year} {2012})}\BibitemShut {NoStop}%
\bibitem [{\citenamefont {Benalcazar}\ \emph {et~al.}(2019)\citenamefont
  {Benalcazar}, \citenamefont {Li},\ and\ \citenamefont
  {Hughes}}]{benalcazar2019}%
  \BibitemOpen
  \bibfield  {author} {\bibinfo {author} {\bibfnamefont {W.~A.}\ \bibnamefont
  {Benalcazar}}, \bibinfo {author} {\bibfnamefont {T.}~\bibnamefont {Li}}, \
  and\ \bibinfo {author} {\bibfnamefont {T.~L.}\ \bibnamefont {Hughes}},\
  }\href {\doibase 10.1103/PhysRevB.99.245151} {\bibfield  {journal} {\bibinfo
  {journal} {Phys. Rev. B}\ }\textbf {\bibinfo {volume} {99}},\ \bibinfo
  {pages} {245151} (\bibinfo {year} {2019})}\BibitemShut {NoStop}%
\bibitem [{\citenamefont {Wheeler}\ \emph {et~al.}(2019)\citenamefont
  {Wheeler}, \citenamefont {Wagner},\ and\ \citenamefont
  {Hughes}}]{wheeler2019}%
  \BibitemOpen
  \bibfield  {author} {\bibinfo {author} {\bibfnamefont {W.~A.}\ \bibnamefont
  {Wheeler}}, \bibinfo {author} {\bibfnamefont {L.~K.}\ \bibnamefont {Wagner}},
  \ and\ \bibinfo {author} {\bibfnamefont {T.~L.}\ \bibnamefont {Hughes}},\
  }\href {\doibase 10.1103/PhysRevB.100.245135} {\bibfield  {journal} {\bibinfo
   {journal} {Phys. Rev. B}\ }\textbf {\bibinfo {volume} {100}},\ \bibinfo
  {pages} {245135} (\bibinfo {year} {2019})}\BibitemShut {NoStop}%
\bibitem [{\citenamefont {Kang}\ \emph {et~al.}(2019)\citenamefont {Kang},
  \citenamefont {Shiozaki},\ and\ \citenamefont {Cho}}]{kang2019}%
  \BibitemOpen
  \bibfield  {author} {\bibinfo {author} {\bibfnamefont {B.}~\bibnamefont
  {Kang}}, \bibinfo {author} {\bibfnamefont {K.}~\bibnamefont {Shiozaki}}, \
  and\ \bibinfo {author} {\bibfnamefont {G.~Y.}\ \bibnamefont {Cho}},\ }\href
  {\doibase 10.1103/PhysRevB.100.245134} {\bibfield  {journal} {\bibinfo
  {journal} {Phys. Rev. B}\ }\textbf {\bibinfo {volume} {100}},\ \bibinfo
  {pages} {245134} (\bibinfo {year} {2019})}\BibitemShut {NoStop}%
\bibitem [{\citenamefont {Jackson}(1999)}]{jackson1999}%
  \BibitemOpen
  \bibfield  {author} {\bibinfo {author} {\bibfnamefont {J.~D.}\ \bibnamefont
  {Jackson}},\ }\href {http://cdsweb.cern.ch/record/490457} {\emph {\bibinfo
  {title} {Classical electrodynamics}}},\ \bibinfo {edition} {3rd}\ ed.\
  (\bibinfo  {publisher} {Wiley},\ \bibinfo {address} {New York, {NY}},\
  \bibinfo {year} {1999})\BibitemShut {NoStop}%
\bibitem [{Note1()}]{Note1}%
  \BibitemOpen
  \bibinfo {note} {It is easy to check that the first (second) moments of the
  charge densities $\rho ({\protect \mathaccentV {vec}17Er})$ and $\rho
  ^\protect \text {macro}({\protect \mathaccentV {vec}17Er})$ are the same if
  the first (first and second) moments of $g({\protect \mathaccentV
  {vec}17Er})$ vanish.}\BibitemShut {Stop}%
\bibitem [{\citenamefont {Vanderbilt}(2018)}]{vanderbilt2018}%
  \BibitemOpen
  \bibfield  {author} {\bibinfo {author} {\bibfnamefont {D.}~\bibnamefont
  {Vanderbilt}},\ }\href
  {https://www.cambridge.org/core/books/berry-phases-in-electronic-structure-theory/DDD71CA4FE9AF5F3A2FB300E602F394A}
  {\emph {\bibinfo {title} {Berry Phases in Electronic Structure Theory:
  Electric Polarization, Orbital Magnetization and Topological Insulators}}}\
  (\bibinfo  {publisher} {Cambridge University Press},\ \bibinfo {year}
  {2018})\BibitemShut {NoStop}%
\bibitem [{\citenamefont {Rhim}\ \emph {et~al.}(2017)\citenamefont {Rhim},
  \citenamefont {Behrends},\ and\ \citenamefont {Bardarson}}]{rhim2017}%
  \BibitemOpen
  \bibfield  {author} {\bibinfo {author} {\bibfnamefont {J.-W.}\ \bibnamefont
  {Rhim}}, \bibinfo {author} {\bibfnamefont {J.}~\bibnamefont {Behrends}}, \
  and\ \bibinfo {author} {\bibfnamefont {J.~H.}\ \bibnamefont {Bardarson}},\
  }\href {\doibase 10.1103/PhysRevB.95.035421} {\bibfield  {journal} {\bibinfo
  {journal} {Phys. Rev. B}\ }\textbf {\bibinfo {volume} {95}},\ \bibinfo
  {pages} {035421} (\bibinfo {year} {2017})}\BibitemShut {NoStop}%
\bibitem [{Note2()}]{Note2}%
  \BibitemOpen
  \bibinfo {note} {In the same spirit, the corner charges and the edge dipoles
  are finer (higher-order) observables compared to the edge charge
  density.}\BibitemShut {Stop}%
\bibitem [{\citenamefont {Souza}\ \emph {et~al.}(2000)\citenamefont {Souza},
  \citenamefont {Wilkens},\ and\ \citenamefont {Martin}}]{souza2000}%
  \BibitemOpen
  \bibfield  {author} {\bibinfo {author} {\bibfnamefont {I.}~\bibnamefont
  {Souza}}, \bibinfo {author} {\bibfnamefont {T.}~\bibnamefont {Wilkens}}, \
  and\ \bibinfo {author} {\bibfnamefont {R.~M.}\ \bibnamefont {Martin}},\
  }\href {\doibase 10.1103/PhysRevB.62.1666} {\bibfield  {journal} {\bibinfo
  {journal} {Phys. Rev. B}\ }\textbf {\bibinfo {volume} {62}},\ \bibinfo
  {pages} {1666} (\bibinfo {year} {2000})}\BibitemShut {NoStop}%
\bibitem [{Note3()}]{Note3}%
  \BibitemOpen
  \bibinfo {note} {\protect \url
  {https://sites.google.com/g.ecc.u-tokyo.ac.jp/workshop-multipole/}}\BibitemShut
  {NoStop}%
\end{thebibliography}%
\end{document}